\documentclass[sigconf]{acmart}
\AtBeginDocument{%
  }

\usepackage{caption}
\usepackage{subcaption}
\copyrightyear{2023} 
\acmYear{2023} 
\setcopyright{rightsretained} 
\acmConference[CHI '23]{Proceedings of the 2023 CHI Conference on Human Factors in Computing Systems}{April 23--28, 2023}{Hamburg, Germany}
\acmBooktitle{Proceedings of the 2023 CHI Conference on Human Factors in Computing Systems (CHI '23), April 23--28, 2023, Hamburg, Germany}\acmDOI{10.1145/3544548.3581252}
\acmISBN{978-1-4503-9421-5/23/04}

\newcommand*{\tl}{}
\newcommand*{\zx}{}

\begin{document}

\title[Inform the uninformed]{Inform the uninformed: Improving Online Informed Consent Reading with an AI-Powered Chatbot}

\author{Ziang Xiao}
\email{ziang.xiao@jhu.edu}
\affiliation{%
    \institution{Johns Hopkins University}
    \city{Baltimore}
    \country{United States}
}
\affiliation{%
    \institution{Microsoft Research}
    \city{Montreal}
    \country{Canada}
}

\author{Tiffany Wenting Li}
\email{wenting7@illinois.edu}
\affiliation{%
    \institution{University of Illinois Urbana-Champaign}
    \city{Urbana}
    \country{United States}
}

\author{Karrie Karahalios}
\email{kkarahal@illinois.edu}
\affiliation{%
    \institution{University of Illinois Urbana-Champaign}
    \city{Urbana}
    \country{United States}
}

\author{Hari Sundaram}
\email{hs1@illinois.edu}
\affiliation{%
    \institution{University of Illinois Urbana-Champaign}
    \city{Urbana}
    \country{United States}
}

\renewcommand{\shortauthors}{Xiao et al.}

\begin{abstract}
 \zx{Informed consent is a core cornerstone of ethics in human subject research. Through the informed consent process, participants learn about the study procedure, benefits, risks, and more to make an informed decision. However, recent studies showed that current practices might lead to uninformed decisions and expose participants to unknown risks, especially in online studies. Without the researcher's presence and guidance, online participants must read a lengthy form on their own with no answers to their questions. In this paper, we examined the role of an AI-powered chatbot in improving informed consent online. By comparing the chatbot with form-based interaction, we found the chatbot improved consent form reading, promoted participants’ feelings of agency, and closed the power gap between the participant and the researcher. Our exploratory analysis further revealed the altered power dynamic might eventually benefit study response quality. We discussed design implications for creating AI-powered chatbots to offer effective informed consent in broader settings.}
\end{abstract}



\maketitle

\section{Introduction}
As a core cornerstone of ethics in human subject research~\cite{faden1986history}, informed consent is the process that guards prospective participants’ voluntary and informed participation decisions. Through the informed consent process, the participant understands the purpose of the study, procedures to be carried out, potential risks and benefits of participation, the extent of data collection and confidentiality, and their rights. Despite its importance, studies found people often sign the form without a thorough read~\cite{lavelle1993factors,sherlock2014patients,moreira2016informed,varnhagen2005informed, pedersen2011undergraduate,ferreira2018informed}. In this study, we examined the role of an AI-powered chatbot in improving the online informed consent process.

Consent form reading research, from high-stake studies with severe ramifications to low-stake studies with minimal risk, shows that participants do not thoroughly read consent forms~\cite{lavelle1993factors,sherlock2014patients,moreira2016informed,varnhagen2005informed, pedersen2011undergraduate,ferreira2018informed}. The participants become less informed when the consent process moves online~\cite{perrault2016informed,geier2021informed,pedersen2011undergraduate}. For example, Ghandour et al.~\cite{ghandour2013giving} found that 65\% people spent less than 30 seconds reading an online consent form with over 800 words, which should have taken roughly seven minutes. Pedersen et al.~\cite{pedersen2011undergraduate} showed that compared to the in-person informed consent process, participants' ability to recognize and recall information from the consent form dipped further when the same consent form was presented online. As a result, the current informed consent process may be neither informed nor consensual.

Uninformed participation decisions put both the participant and the researcher at risk. \zx{The consequence could endanger the participant's health if they failed to notice the study procedure could induce stress on their pre-existing mental or physical conditions~\cite{del2005informed}.} It could also create privacy risks to participants' personal data if the participant holds incorrect assumptions of the researcher's data-sharing practice~\cite{cummings2015impact}. Further, a lack of a good understanding of the consent form may inhibit the participant's autonomy in making the consent decision. Cassileth et al. ~\cite{cassileth1980informed} found more than 1/4 of participants in a medical study thought accepting the consent form was the only option to receive treatment even though the form indicated alternatives. Meanwhile, an ineffective informed consent process could damage study validity and data quality ~\cite{douglas2021some,heider2020perceptions}. Failure to understand the study procedures and purposes may not only negatively impact performance on experimental tasks but also create confounding factors, especially for studies involving deception ~\cite{douglas2021some,heider2020perceptions}. Conversely, an effective informed consent process could improve participant engagement and promote trust and rapport between the participant and the researcher~\cite{rowbotham2013interactive, douglas2021some,halpern2002continuing}. It could reduce the power asymmetry in participant-researcher relations by bridging the information gap of the study, informing participants of their rights, and guarding voluntary decisions ~\cite{schuck1994rethinking}. Consequently, a successful informed consent process benefits both the participant and the researcher. 

In recent years, especially during the COVID-19 pandemic, online informed consent has become more prevalent. More studies, both online and in-person, collected participants' consent remotely. And those studies may involve risky procedures or collect sensitive information, including people's mental health and personal genetic data. Therefore, improving online informed consent reading is becoming increasingly important.

However, improving consent-form reading is a challenging task, especially in an online environment~\cite{assale2019digitizing,flory2004interventions,pedersen2011undergraduate}. Compared to an in-person setting where the researcher could directly interact with the participants, guide them through the consent form, and answer their questions, in an online environment, the absence of the researcher creates communication barriers and significantly demotivates and disincentivizes consent form reading~\cite{douglas2021some,perrault2016informed,walkup2009prospective}. In the past, researchers have experimented with different interventions to improve consent form reading, including simplifying the content, lowering reading grade level, and introducing interactive features~\cite{perrault2016informed, anderson2017improving, perrault2019concise}. However, two literature reviews of over 70 past studies suggest that the effect of those interventions was limited, and it is burdensome to design and develop compelling and effective interactive experiences~\cite{flory2004interventions,de2020implementation}. Therefore, exploring new techniques to improve online consent form reading is necessary.

\zx{We examined the role of an AI-powered chatbot \footnote{In contrast to traditional rule-based chatbots, we define an AI-powered chatbot as a chatbot that leverages artificial intelligence (AI) technologies, including machine learning, natural language processing, and advanced analytics in the building process or the deployment environment of the chatbot.} in the delivery of consent form content. We built the chatbot {\tt Rumi} with a hybrid system that combines a rule-based system with AI components. {\tt Rumi} can greet a participant, go through the consent form section by section, answer the participant's questions, and collect people's consent responses, similar to what an experienced researcher would do in a lab setting. Through a conversational interface, a chatbot could grab people's attention, deliver personalized experiences, and provide human-like interactions~\cite{xiao2020tell}. All those features could potentially benefit the online informed consent process. However, chatbots also bear several risks. First, a turn-by-turn chat requires extra time and effort to complete the informed consent process, which is a major challenge in consent form reading~\cite{flory2004interventions}. The risk is even higher for studies with paid participants, who would not be rewarded for taking longer to complete the study. Second, current chatbots are far from perfect. Their limited conversation capabilities may deliver incorrect answers or lead to user disappointment and frustration~\cite{grudin2019chatbots}. Therefore, it is yet unknown how a chatbot could affect the online informed consent process.}

To explore the effectiveness of an AI-powered chatbot that guides a participant through an informed consent process, we asked three research questions (RQs),

\textbf{RQ1:} \zx{How would the participant's consent form reading differ in a study with the AI-powered chatbot-driven consent process vs. the form-based consent process? \textit{(Consent Form Reading)}}
        
\textbf{RQ2:} \zx{How would the participant's power relation with the researcher differ in a study with the AI-powered chatbot-driven consent process vs. the form-based consent process? \textit{(Power Relationships)}}
        
\textbf{RQ3:} \zx{How would response quality differ in a study with the AI-powered chatbot-driven consent process vs. the form-based consent process? (\textit{Study Response Quality})}
    
To answer our research questions, we designed and conducted a between-subject study that compared the use of an AI-powered chatbot, {\tt Rumi}, and a typical form-based informed consent process in an online survey study. Since no previous work has examined the use of chatbots that deliver informed consent, in this study, we focused on examining the holistic effect of a chatbot instead of the effect of individual chatbot features.  With a detailed analysis of 238 study participants' informed consent experiences and their responses to the survey study, we found 1) {\tt Rumi} improved consent form reading, in terms of both recall and comprehension, 2) participants who completed the consent form with {\tt Rumi} perceived a more equal power-relation with the researcher, 3) by reducing the power gap, the improved informed consent experience ultimately benefited study response quality. 

To the best of our knowledge, our work is the first that systematically compared the \textit{holistic} effect of an AI-powered chatbot-driven informed consent process with that of a typical online informed consent process. Our work provides three unique contributions.

\begin{enumerate}
    \item An understanding of the holistic effect of AI-powered chatbots conducting online informed consent. Our findings extend prior work on creating an effective informed consent process and reveal the practical value of using an AI-powered chatbot for informed consent, especially in improving consent form recall and understanding, researcher-participant relation, and study response quality. We further provided empirical evidence that could attribute observed improvement in study response quality to the reduced power gap.
    
    \item Design implications of creating effective AI-powered chatbots for informed consent. We discussed design considerations, such as personalized reading assistance and power dynamics management, to further improve the online informed consent process.
    
    \item New opportunities for creating and operationalizing AI-enabled, consent experiences for a broader context. The demonstrated effectiveness of the chatbot-driven informed consent opens up opportunities for employing AI-powered chatbots for other types of consent in the age of AI.
\end{enumerate}
\section{Related Work}
\subsection{Consent Form Reading}
The concept of informed consent is embedded in the principles of many ethical guidelines including the Nuremberg Code, The Declaration of Helsinki, and The Belmont Report~\cite{faden1986history}. Four core elements ensemble the informed consent process, including disclosure, comprehension, voluntariness, and competency. Through the informed consent process, participants will learn about the study's purpose, procedure, risks, and benefits to make an informed decision. 

Despite its importance, people often sign the consent form without a thorough read, regardless of whether it is about a clinical trial that may risk their physical and mental health or a study about their political opinions. For example, Lavelle-Jones et al.~\cite{lavelle1993factors} found that 69\% of patients preparing to undergo various surgical procedures signed the consent form without reading it carefully. Varnhagen et al.~\cite{varnhagen2005informed} showed that in a study regarding technology use, study participants could only recall less than 10\% of the information contained in the form, and 35\% of participants reported that they only skimmed the form or did not read it at all. Cummings et al. attributed the discrepancy between people's study participation decisions and their concerns regarding confidentiality, anonymity, data security, and study sensitivity in studies with open-data sharing practices to participants' inattentiveness to the consent form~\cite{cummings2015impact}. Consent form reading has become increasingly challenging as it moves online~\cite{perrault2016informed,geier2021informed,pedersen2011undergraduate}. An online consent form is low-cost and easy to administrate, in addition to its broaden-reach across the internet. However, compared to the in-lab setting, no researcher will guide the participant through the consent form, explain the content, and clarify the participant's question~\cite{flory2004interventions,de2020implementation,anderson2017improving}. Therefore, online informed consent often yields less informed participation decisions, especially when the study is more complicated and riskier.

Ineffective informed consent reading not only puts participants under unaware risks but also harms the study's validity and data quality~\cite{rowbotham2013interactive, douglas2021some,halpern2002continuing}. Consent forms contain important information about study procedures and purposes, and comprehending such information may determine the success of later study manipulation, especially for deception studies with cover stories. Unaware confounding factors may further contribute to the replication crisis in social sciences. 

Altogether, while the informed consent process plays a vital role for both participants and researchers in a research study, the current practice of conducting an informed consent process has many weaknesses, especially since more studies have started to collect participants' consent online. In this study, we aim to improve the online consent form reading by exploring novel interaction techniques, e.g., an AI-powered chatbot. 

\subsection{Improving Consent Form Reading}
While many prior studies have focused on improving consent form reading, the study results were not always consistent~\cite{flory2004interventions,de2020implementation}. One group of researchers focused on the design of the consent form, including text readability~\cite{morrow1980readable,gray1978research,bjorn1999can}, length~\cite{dresden2001modifying,perrault2019concise,perrault2016informed}, layout~\cite{coyne2003randomized}, and media~\cite{agre2003improving,friedlander2011novel,tait2014enhancing}. Dresden and Levitt~\cite{dresden2001modifying} found study participants could retain more information from a consent form with less unnecessary information and simpler vocabulary. However, other studies found a concise consent form may not yield a higher comprehension score~\cite{perrault2019concise,bjorn1999can}. Although study participants advocate for a shorter form with simpler language, due to the regulation and the nature of a study, it is often difficult to achieve~\cite{geier2021informed}. Researchers also explored converting text-based forms into multimedia. For example, Friedlander et al. found the utility of using video to deliver a consent form in increasing people's engagement and comprehension~\cite{friedlander2011novel}. However, there is no conclusive evidence of its effectiveness according to multiple meta-analyses~\cite{flory2004interventions,palmer2012effectiveness}. 

Another group of researchers brought interactivity into the informed consent process to create an engaging and personalized experience that facilitates consent form reading \cite{bickmore2015automated,balestra2016effect,anderson2017improving}. In an in-person setting, letting the researcher go through the consent form and answer the participant's questions is deemed the most effective and desirable~\cite{anderson2017improving}. As more studies move online, researchers have started to explore new interactive features to improve online consent~\cite{balestra2016effect, bickmore2015automated}. One effective intervention is to test the participant's knowledge about the consent form before signing~\cite{knepp2018using}. Further, Balestra et al. used social annotations in online consent forms to facilitate consent form comprehension~\cite{balestra2016effect}. Bickmore et al. built an embodied agent that can explain a medical consent form to the reader and found people to be more satisfied with the agent that can tailor its explanation to the participant's existing knowledge~\cite{bickmore2015automated}. Although not all attempts were successful, many studies emphasized the importance of interaction with the researcher to ensure participants' understanding of the consent information and to foster trust, especially for complex and risky studies~\cite{chen2020replacing}. 

\zx{We built {\tt Rumi} with a state-of-art hybrid chatbot framework with AI-powered question-answering modules. Compared to prior work~\cite{bickmore2015automated,zhou2014agent} where the agent largely relies on a rule-based system without taking natural language input, recent language technologies enabled {\tt Rumi} to answer a diverse set of questions in natural language and deliver engaging experiences with multiple conversational skills (for more details, see Sec. \ref{sec:system}). Such a chatbot opens up a new opportunity to bring humanness back to the informed consent process and make the informed consent process more effective.}

\subsection{Power Relation in Human Subject Research}
In a power relation, the person with lower social-formative power is often constrained by their superior counterpart~\cite{foucault1982subject}. In the context of human subject research, the researcher has often been viewed as having total authority and being able to decide the resource distribution whereas the participant is often sitting at the lower end~\cite{karnieli2009power}. The power asymmetry between the researcher and the participant results from the researcher's control over the participant's recruitment, treatment, data, and compensation~\cite{cassell1980ethical}. Such a power gap inhibits the participant's autonomy, decreases study engagement, and deters authentic answers~\cite{cassell1980ethical,karnieli2009power}. 

Many researchers advocate for power redistribution to close the power gap in human subject research for both ethical and data quality considerations~\cite{ebbs1996qualitative, wolf2018situating}. By reducing the power gap, the participants could be more engaged during the study, more comfortable disclosing their true thoughts, and more cooperative with the study procedure, which may ultimately benefit the study quality. For example, Chen analyzed the interviewer's language use and found the reduced power gap encourages data richness~\cite{Chen2011-el}. However, some studies warn of the importance of maintaining the distance between the researcher and the participant for professional judgment~\cite{lincoln1985naturalistic,torres2002evolving}. 

By sharing information about the study procedure, clarifying risks and benefits, and elaborating on the participant's rights, the informed consent process is designed to close the information gap and ensure the participant's autonomy~\cite{schuck1994rethinking}. \zx{Additionally, as one of the earliest interactions happens between the researcher and the participants, the informed consent process provides an excellent opportunity to redistribute power and establish trust.} Kustatscher \cite{kustatscher2014informed} used visual aids to improve the informed consent for children and found such an engaging process altered the power relation and created a more comfortable environment for information disclosure.

We further extended prior knowledge on the power relation in the research setting by showing the improved informed consent process with an AI-powered chatbot could close the power gap. Through our path analysis, we found the observed effect on study quality can be attributed to the altered power relation.

\subsection{Conversational AI as a Research Tool}
Recent advances in natural language processing enabled more powerful chatbots for researchers. An emerging application is using chatbots as a tool for research success, including in-depth conversational surveys online~\cite{devault2014simsensei,xiao2020tell}, or field studies in the real world~\cite{williams2018supporting,tallyn2018ethnobot,kim2019comparing}. Compared to traditional form-based interactions, a chatbot retains the advantage of scalability while providing more engaging and personalized experiences~\cite{xiao2020tell}. Specifically, the conversational interface provides interactivity through a turn-by-turn chat, which allows a chatbot to frame questions in a more personalized, conversational way, deliver human-like social interactions to encourage self-disclosure, and probe for in-depth information~\cite{xiao2020if}. Tallyn et al.~\cite{tallyn2018ethnobot} demonstrated the effectiveness of using a chatbot to gather ethnographic data in the absence of a human ethnographer. Xiao et al. showed that compared to a form-based survey, an AI-powered chatbot could manage a conversational survey, collect higher quality data, and deliver a more engaging survey experience~\cite{xiao2020tell}. Chatbots have also been used for other research scenarios, including building psychological profiles~\cite{zhou2019trusting}, delivering interventions~\cite{fadhil2017addressing}, and instructing study procedures~\cite{kumar2016sanative}. For instance, psychologists used a chatbot to infer a participant's personality through conversation to avoid faking in questionnaire-based assessment~\cite{zhou2019trusting,xiao2019should}. Lee et al. built chatbots to practice journaling in the context of mental health research~\cite{lee2020hear}. In summary, past studies have shown a chatbot can potentially serve as a moderator like a research assistant or an interviewer to proactively manage a study process to collect high-quality data for research success.

Our study examined the utility of chatbots in another critical step in human subject research, the informed consent process. We demonstrated a new opportunity and provided design implications to streamline the use of conversational AI in research. We offered design implications to build better future AI for social science.  
\section{Method}
To answer our RQs, we designed a between-subject study that compared the outcomes of two methods to deliver the consent form online, an AI-powered chatbot (\textit{Chatbot Condition}) and a typical form (\textit{Form Condition}), on consent form reading, participant-researcher power relation, and study response quality.

\subsection{Dummy Survey Study Design}
To understand how an AI-powered chatbot facilitates online informed consent reading and how it might influence study response quality, we have to separate the study quality evaluation from the consent form reading evaluation. Therefore, we designed a dummy study dedicated to evaluating study response quality. 

The dummy study is about problematic social media use. To complete the study, the participant first read a short article about problematic social media use. The goal is to familiarize participants with the issue. Then, the participant answered a survey with both close-ended questions and open-ended questions. The choice-based questions included six five-point Likert Scale questions adopted from Bergen Social Media Addiction Scale~\cite{lin2017psychometric}. The open-ended questions are adopted from~\cite{Baylor_University2016-qa} with the goal of understanding people's attachment to social media and how it affects people's real life. Both question sets are widely used in social media research. We choose this topic for four reasons. First, it relates to most people online and is suitable to conduct online. Second, we could vary the level of psychological discomfort and data sensitivity to simulate a wider range of online studies. \zx{Third, the survey method is the most widely used research instrument, ensuring our finding's generalizability.} Lastly, prior studies have provided us with established methods to robustly measure the response quality of a survey with both open-ended questions and choice-based questions~\cite{xiao2020tell,mulligan2001nondifferentiation}. We also considered a genetic study used by~\cite{balestra2016effect} that deals with genetic data. Although mishandled genetic data may cause more tangible consequences, the level of risks is difficult to vary and the generalizability is limited.

We designed three versions of our dummy study for three common risk levels in online survey studies,

\begin{itemize}
    \item \textit{Low} -  Non-sensitive data without personal identifiers
	\item \textit{Medium} - Sensitive data without personal identifiers
	\item \textit{High} - Sensitive data with personal identifiers
\end{itemize}

\zx{For the \textit{Low} risk version, the survey will ask about people's \textit{opinions} regarding others' problematic social media use. This version is designed to evoke minimal psychological discomfort by asking for opinions about other people instead of directly recalling their own experiences.} For the \textit{Medium} risk version, the participant will answer questions regarding \textit{their own} problematic social media use. And in the \textit{High} risk version, we will ask participants to additionally reveal their social media handles as a personal identifier for a follow-up study. The distinction was made clear in the study description and potential risks in the consent form. 

\subsection{Consent Form Design: Form Condition}
The consent form was based on the Social Behavioral Research Consent Form template provided by the Institutional Review Board (IRB) at the University of Illinois Urbana-Champaign. We improved the design of the form-based consent form based on recommendations from prior work~\cite{dresden2001modifying,perrault2019concise,perrault2016informed,coyne2003randomized}. Specifically, we broke the consent form into sections to reduce information overload and used a clear and simple format to ensure clarity and readability. 

\subsection{Consent Form Design: Chatbot Condition}
\label{sec:system}
We created a chatbot, {\tt Rumi}, with the goal of simulating an in-person informed consent process experience where a researcher goes through the consent form section by section, asks if the participant has questions, and makes clarifications. In our study, {\tt Rumi} first greeted the participant and informed the participant that it could take questions at any time during the informed consent process. Then, {\tt Rumi} went through the informed consent form section by section with the exact content in the Form Condition. Participants can click the ``Next'' button to proceed to the next section or type in the text box with their questions or other requests. During the process, {\tt Rumi} proactively asked if the participant had any questions twice, one after the risk section and one at the very end. Participants can skip by pressing the ``No Questions'' button. Then, {\tt Rumi} confirmed the participant's age and elicited for their consent to join the study. \zx{We included a video in the supplementary material to demonstrate how {\tt Rumi} conducts the informed consent process.}

\zx{We adopted a hybrid approach to build {\tt Rumi} by combining a rule-based model with AI-powered modules~\cite{xiao2023powering}. We made this decision for the following reasons. First, rule-based models have limited capability to recognize participants' questions and deliver diverse and engaging responses, let alone handle non-linear conversations (e.g., side-talk, request to repeat information, etc.). Second, although an end-to-end generative system could lead to a more engaging experience, it may produce incorrect or non-factual answers. Those answers may cause severe issues in high-stake scenarios~\cite{bickmore2018patient,blodgett2020language}, e.g., delivering a consent form. Third, the consent form content is typically specific to the situation or context in which the form is being used. Such a few-shot or zero-shot setting poses another challenge to building an effective generative model, even fine-tuning pre-trained models (e.g., DialoGPT~\cite{zhang2019dialogpt}). As a result, we decide to prioritize the answer authenticity and build {\tt Rumi} on a hybrid system with AI-powered modules to enable {\tt Rumi}'s ability to handle a broad set of questions and provide diverse and accurate answers.}
 
\zx{Specifically, we built {\tt Rumi} on the Juji platform, a hybrid chatbot building platform. Juji provides built-in AI-based functions for dialog management and effective Q\&A~\cite{jujidoc}. {\tt Rumi} will follow a rule-based conversation agenda to go through the consent form section by section and acquire the participant's consent. Given a question, Juji offers pre-trained Natural Language Understanding (NLU) models to identify relevant questions with known answers in a Q\&A database and returns an answer or a follow-up question for clarification. When the chatbot is unsure about how to answer a question, it will recommend similar questions to give participants a chance to obtain desired answers and to learn more about the chatbot's capabilities~\cite{jujidoc}. Juji also offers a diverse set of conversation skills such as handling side talks and conversation repair to provide an engaging conversation experience~\cite{xiao2020tell}.}

We curated the Q\&A database by creating a set of seed questions ourselves and piloting {\tt Rumi} with 54 online participants. Since the goal is to gauge potential questions participants may ask, we asked our participants to ask as many questions as possible. Researchers on the team wrote answers for each question based on the consent form and added Q\&A pairs into the database. Before deploying {\tt Rumi} for this study, the Q\&A database contains over 200 Q\&A pairs. To further enhance {\tt Rumi} ability to recognize questions that may be differently phrased, we leveraged a text-generation model GPT-3 (text-davinci-002)~\cite{ouyang2022training} to create question paraphrases. Similarly, we used the GPT-3 to create a candidate answer set for each question. We augmented the Q\&A database with five question paraphrases and five candidate answers for each Q\&A pair. We hand-checked all generated texts to ensure information authenticity.

\begin{figure*}[t]
    \centering
    \includegraphics[width=0.90\textwidth]{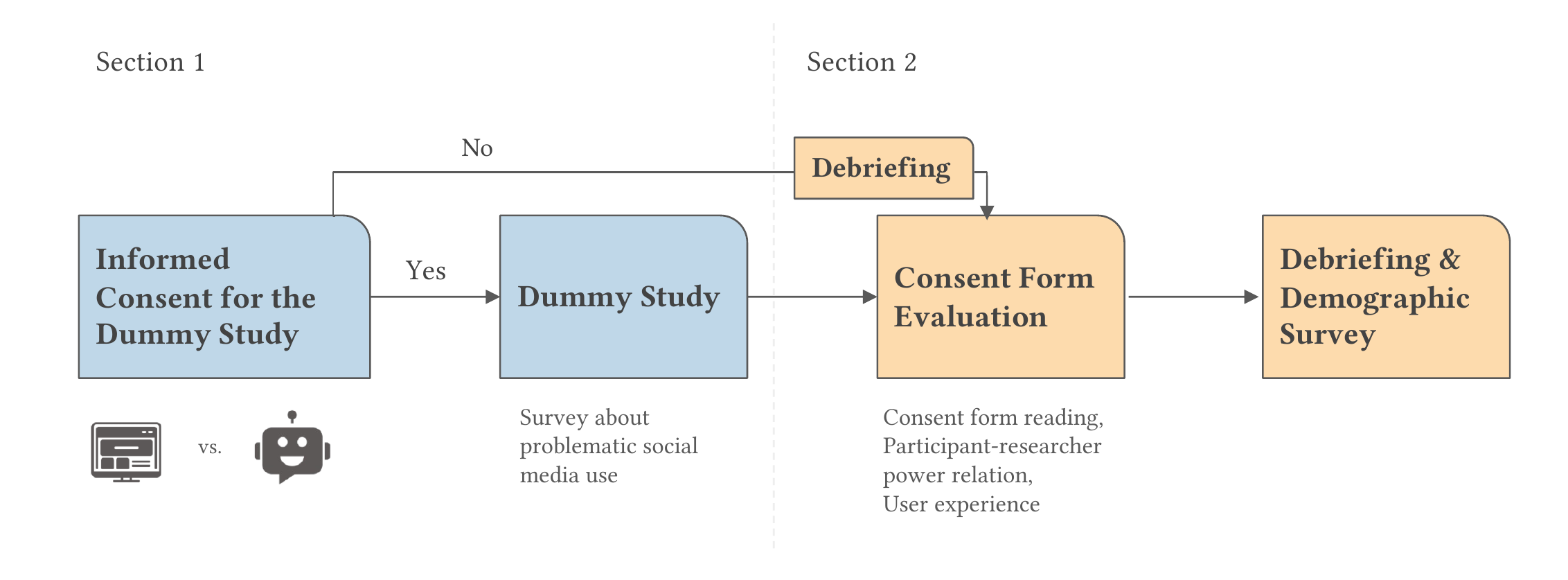}
    \caption{\small The figure shows the overall study procedure. In Section 1, based on the assigned condition, the participant interacted with {\tt Rumi} or the Form for the informed consent process and then completed a dummy study about social media use. In Section 2, participants answered questions about their consent form reading experience. We debriefed our participants and collected demographic information at the end. If the participants decided not to join the study in Section 1, they would be invited to start Section 2.}
    \label{fig:study procedure}
\end{figure*}

\subsection{Study Procedure}
The complete study consists of two sections (Fig~\ref{fig:study procedure}) and is approved by the IRB. In the first section, participants were randomly assigned to one of the conditions (Chatbot Condition vs. Form Condition) and one of the risk levels (Low vs. Medium vs. High). The study started with the informed consent process based on the assigned condition. Upon consent, the participant completed the Dummy Survey Study. 

Our consent form evaluation is in the second section. The participants answered questions about their understanding of the consent form, their perceived relationship with the researcher, the informed consent process experience, and demographic information. In the end, we debriefed the participants with the real purpose of the study and asked them to complete an additional consent form about sharing their answers to the second section of the study. Before leaving the study, we also asked participants to give open-ended comments on their informed consent. 

If the participant decided not to join the study in the first section, they were debriefed on the real purpose of the study and asked if they were willing to join our second section to evaluate their consent form reading. If they agreed to join, we would direct them directly to section two to complete the rest of the study. 

\subsection{Measures}

\subsubsection{\textbf{Consent Form Reading}} We measure participants' consent form reading in two dimensions, \textit{recall} and \textit{comprehension}. 

\textit{Recall:} The ability to recall information from the consent form suggests that people pay attention to the consent form. Similar to ~\cite{douglas2021some}, to assess participants' recall of the form, we inserted two random statements ``\textit{watch a video with orange cat}'' and ``\textit{read study materials in a blue background}'' into the middle of the study procedure section and the risks section of the consent form respectively. Participants were tested on their recall of the color words, ``\textit{orange}'' and ``\textit{blue}''. We decided to use recall of two random statements as our measure of consent form reading because it could sensitively measure thoroughness and making a correct guess is difficult~\cite{douglas2021some}. Studies showed participants were more likely to read the procedure and risks sections carefully \cite{varnhagen2005informed,douglas2021some}, which provides us an opportunity to measure the lower bound of consent form reading. The participants were asked to select one or two color words from five common color words (``green'', ``red'', ``white'', ``orange'' and ``blue''). The participants would receive a score of 2 if both color words were selected correctly, a score of 1 if the participants selected only one option and the answer was correct, and a score of 0.5 if the participant selected two options but was partially correct. Otherwise, they got a 0. 

\textit{Comprehension:} Participants' comprehension of the consent form reflects the effectiveness of reading. To comprehend the consent form, the participants need to process the text, understand its meaning, and integrate it with their prior knowledge \cite{grabe2008reading}. Inspired by \cite{grabe2008reading}, we measured comprehension with six questions that required participants to process the information presented in the consent form beyond simple recollection. These questions asked about the study procedure, potential risks, and actions to take if a certain scenario happens, e.g., how the participants could protect their privacy if a data breach happens or what a participant should do if they decide to withdraw from the study. A total of four multiple-choice questions were presented. The comprehension score measured how accurately a participant answered those questions. The final score is the percentage of correct answers, ranging from $0\%$ to $100\%$.

\subsubsection{\textbf{Participant-Researcher Power Relation}} 
We measure the participant-researcher power relation by participants' perceived relationship with the researcher in the study and their feeling of agency and control. 

We measure two aspects of the perceived relationship, Partnership and Trust. Based on \cite{henderson2003power}, we asked if the participants perceived their relationship with the researcher who ran the study as partners. Adopted from \cite{balestra2016effect}, we measured the trust by asking to what degree the participants trusted the researcher would handle their data properly.

According to \cite{cassell1980ethical,karnieli2009power}, the power asymmetry between research parties inhibits participants' autonomy. By power redistribution, we would expect participants to regain agency and control, two constructs of autonomy. We measured the perceived agency and control from both sense of positive agency and the sense of negative agency. We adapted scales from \cite{tapal2017sense} based on the context of the informed consent process. All items adopted a 7-point Likert scale from Strongly Disagree to Strongly Agree.

\subsubsection{\textbf{Study Response Quality}}
We measured the study response quality by examining participants' responses to both choice-based questions and open-ended questions in the problematic social media use survey. 

\textit{Non-differentiation:} Non-differentiation is a survey satisficing behavior where the respondents fail to differentiate between the items by giving nearly identical responses to all items using the same response scale \cite{kim2019straightlining}. Non-differentiation deteriorates both the reliability and validity of question responses. It further inflates intercorrelation among items within the questionnaire and suppresses differences between the items \cite{yan2008nondifferentiation}. We used the mean root of pairs method \cite{mulligan2001nondifferentiation} to measure the non-differentiation in choice-based questions. We calculated the mean of the root of the absolute differences between all pairs of items in a questionnaire. The metric ranged from 0 (The least non-differentiation) to 1 (The most non-differentiation).

\textit{Response Quality Index (RQI):} To measure the response quality of open-ended questions, we created a \textit{Response Quality Index} based on \cite{xiao2020tell}. It measures the overall response quality of N responses given by a participant on three dimensions, relevance, specificity, and clarity, derived from Gricean Maxim \cite{grice1975logic}:  
\begin{equation}
\begin{array}{l}
    \text{RQI} = \sum_{n=1}^{N}  \text{relevance}[i] \times  \text{clarity}[i] \times  \text{specificity}[i] \\ 
   \text{(N is the number of responses in a completed survey)}
   \end{array}
\label{equ:rqi}
\end{equation} 

\textit{Relevance.} A good response should be relevant to the context. For an open-ended question, a quality response should be relevant to the survey question. Irrelevant responses not only provide no new information but also burden the analysis process. We manually assessed the relevance of each open-text response on three levels: 0 – Irrelevant, 1 – Somewhat Relevant, and 2 – Relevant.

\textit{Specificity.} Quality communication is often rich in details. Specific responses provide sufficient details, which help information collectors better understand and utilize the responses and enable them to acquire more valuable, in-depth insights. We manually assessed the specificity of each open-text response on three levels: 0 – Generic description only, 1 – Specific concepts, and 2 – Specific concepts with detailed examples.

\textit{Clarity.} Clarity is another important axis. Each text response should be easily understood by humans without ambiguity, regardless of its topical focus. We manually scored each free-text response on three levels: 0 – Illegible text, 1 – Incomplete sentences, and 2 – Clearly articulated response. 

\textit{Coding Process:} We coded a total of 1428 open-ended responses. \zx{Two researchers with a background in human-computer interaction and expertise in open-ended survey analysis conducted the coding process. They first randomly selected 10\% of the responses and created a codebook on the above three dimensions with definitions and examples. Then, two researchers coded the rest of the data independently and were blind to the condition. Krippendorff’s alpha ranged from 0.83 to 0.98 for each set of coding. The final disagreement was resolved by discussion.}

\subsubsection{\textbf{Participant Experience and Demographics}} 

\textit{Time and Effort:} Injecting interactivity often means the participants need to spend extra time and effort to interact with the system, which is a major trade-off \cite{flory2004interventions,de2020implementation}. Therefore, to measure the perceived time and effort of the informed consent process, we adapted the ASQ scale with two items \cite{lewis1991psychometric} on how satisfied people were with the time and effort spent on the informed consent process, which we later averaged into a single score.

\textit{Future Use:} People's willingness to use the same system in the future is a strong indicator of good user experience and satisfaction. We used a single-item 7-point Likert scale to ask if the participant would use the chatbot or the form to complete an informed consent process in the future. 

\textit{Demographics:} Prior studies on chatbots suggests individual differences in people's experience \cite{xiao2020tell} may moderate their chatbot experience. We collected basic demographic information, including age, gender, education level, and annual household income.

\subsection{Participant Overview}
We recruited fluent English speakers from the United States on Prolific \footnote{www.prolific.co}. Of the 278 participants who opened our link and started the informed consent process, 252 completed the informed form. Two participants in the Chatbot Condition explicitly declined to join the study and left the study immediately. 

Out of the 250 participants who started the study, 238 (Denoted as P\#) completed the study and passed our attention and duplication check. Our following analysis is based on those 238 valid responses (Table \ref{tab:participants}). Among those 238 participants, 97 identified as women, 136 identified as men, and 5 identified as non-binary or third gender. The median education level was a Bachelor's degree. The median household income was between \$50,000 - \$ 100,000. And the median age of participants was between 25 - 34 years old. On average, our participants spent 1.24 mins (SD = 3.03) completing the informed consent process in the Form Condition and 7.75 mins (SD = 7.06) with {\tt Rumi}. We compensated our participants at the rate of \$12/hr.

\begin{table}[]
\begin{tabular}{lllll}
\hline
                  & Low & Medium & High & Total   \\ \hline
Chatbot Condition & 41  & 38     & 40   & 119 \\
Form Condition    & 39  & 40     & 40   & 119 \\
Total             & 80  & 78     & 80   & 238 \\ \hline
\end{tabular}
\caption{\small The table shows the participant distribution across conditions. Participants were randomly assigned to one condition based on the consent form conditions and the risk level. A total of 238 participants were included in the final analysis. }
\label{tab:participants}
\end{table}

\subsection{\zx{Data Analysis}} \label{data_analysis}
\zx{We used Bayesian analysis to compare the distributions of effects on consent form reading (RQ1), the participant's power relation with the researcher (RQ2), and study response quality (RQ3) between two consent methods. We were motivated to use Bayesian analysis for the following reasons. First, Bayesian models allow us to foreground all aspects of the model; No modeling assumptions need checking that are not already foregrounded in the model description. Second, compared to the null hypothesis significance testing (NHST), the Bayesian analysis focuses on ``how strong the effect size is'' instead of ``if there is an effect''. It better fits the exploratory nature of our study. Third, Bayesian models facilitate the accumulation of knowledge within the research community as study outcomes can be used as informative priors later. Kay et al. provide a detailed review of the Bayesian method's advantages in HCI research \cite{kay2016researcher}.}

\zx{We formulated a hierarchical Bayesian model for each outcome measure. We build two types of hierarchical Bayesian models, linear regression models for continuous measures and ordinal logistic regression for ordinal measures. For Recall, Comprehension, Agency and Control \footnote{\zx{We modeled the perceived Agency and Control as a continuous variable as it is a composite score from an 11-item scale.}}, Non-differentiation, and RQI, we} \tl{modeled the data as a Normal distribution and used linear regression models to estimate the Normal distribution means for both the Chatbot Condition and the Form Condition. By contrasting the posterior distributions of the means for the two conditions, we would know how the consent method affects outcome variables. Furthermore, we estimated the effect size of the difference of the posterior distribution with Cohen's d for Bayesian linear regression models.}

\zx{For Partnership, Trust, Time and Effort, and Future Use, we used ordinal logistic regression models to estimate the posterior distributions of the cumulative odds for a given value on the ordinal scale. Our Bayesian analysis goal was to compare whether the rating distribution was significantly different between conditions with respect to the neutral midpoint of the scale (\textit{Neither agree nor disagree}). This would tell us whether participants were more likely to disagree with the statement in one condition over the other. We constructed the distribution of the difference between the cumulative odds of a rating of 4 or below (the midpoint of a 7-point Likert scale) between the Chatbot Condition and the Form Condition. Negative values indicate that participants in one condition had less odds of providing a neutral or negative response to the other condition. We used this for all the above measures except Recall where we estimated the posterior distributions of the cumulative odds of getting the recall question 50\% correct or lower. Based on the posterior distributions, we calculated the Odds Ratio (OR) as the effect size \footnote{\zx{We interpreted the magnitudes of odds ratios based on Chen et al.~\cite{chen2010big} where OR = 1.68, 3.47, and 6.71 are equivalent to Cohen's d = 0.2 (small), 0.5 (medium), and 0.8 (large), respectively}}.}

In all models, we controlled for the following covariates: study risk levels, participants' age, gender, education level, and annual household income. We controlled for these demographics as prior studies on conversational agents suggest individual differences may affect their interaction with a conversational agent~\cite{xiao2020tell}. Full mathematical descriptions of each type of model are provided in the Supplementary Material. We performed the Bayesian analysis using NumPyro \footnote{\zx{https://num.pyro.ai/}}, a popular Bayesian inference framework. We used Markov Chain Monte Carlo (MCMC), a stochastic sampling technique to sample the posterior distribution P($\theta$|D), the distribution functions of the parameters in the likelihood function given the data observations D. In particular, we used the No-U Turn Sampler (NUTS) for sampling.

We supplemented our quantitative analysis with qualitative evidence by analyzing participants’ chat transcripts in the Chatbot Condition. We performed the thematic analysis~\cite{gibbs2007thematic} on the questions people asked. A member of the research team first performed open coding on the data and then refined these codes in an iterative and reflexive process. The same person then used axial coding to group these codes into larger categories to extract common themes.

\begin{figure*}[]
        \centering
        \begin{subfigure}[t]{0.30\textwidth}
            \centering
            \includegraphics[width=\textwidth]{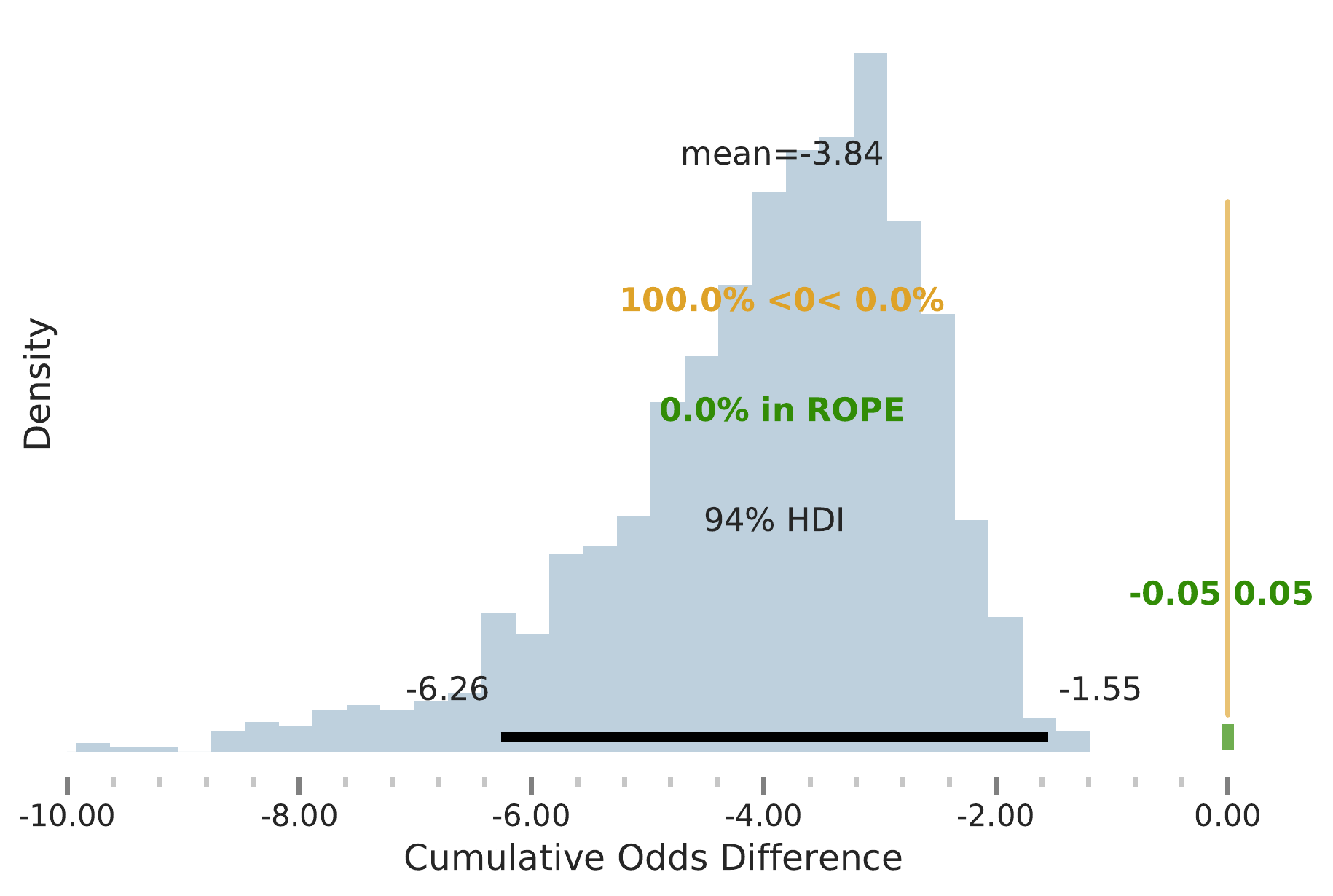}
            \caption[]%
            {{\small Recall, Contrast}}    
            \label{fig:R-contrast}
        \end{subfigure}
        \begin{subfigure}[t]{0.30\textwidth}  
            \centering 
            \includegraphics[width=\textwidth]{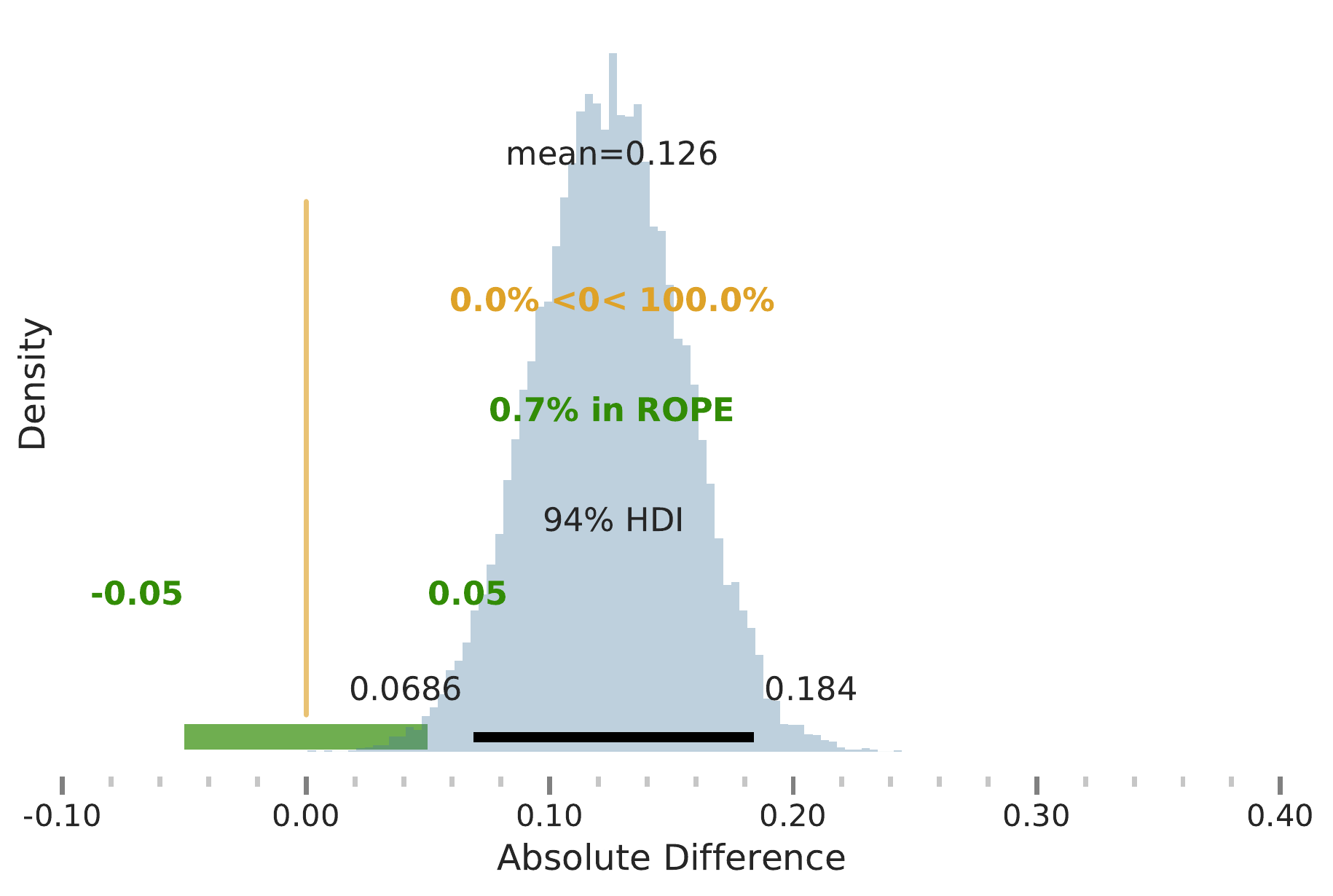}
            \caption[]%
            {{\small Comprehension, Contrast}}    
            \label{fig:C-contrast}
        \end{subfigure}
        \vskip\baselineskip
        \begin{subfigure}[b]{0.30\textwidth}   
            \centering 
            \includegraphics[width=\textwidth]{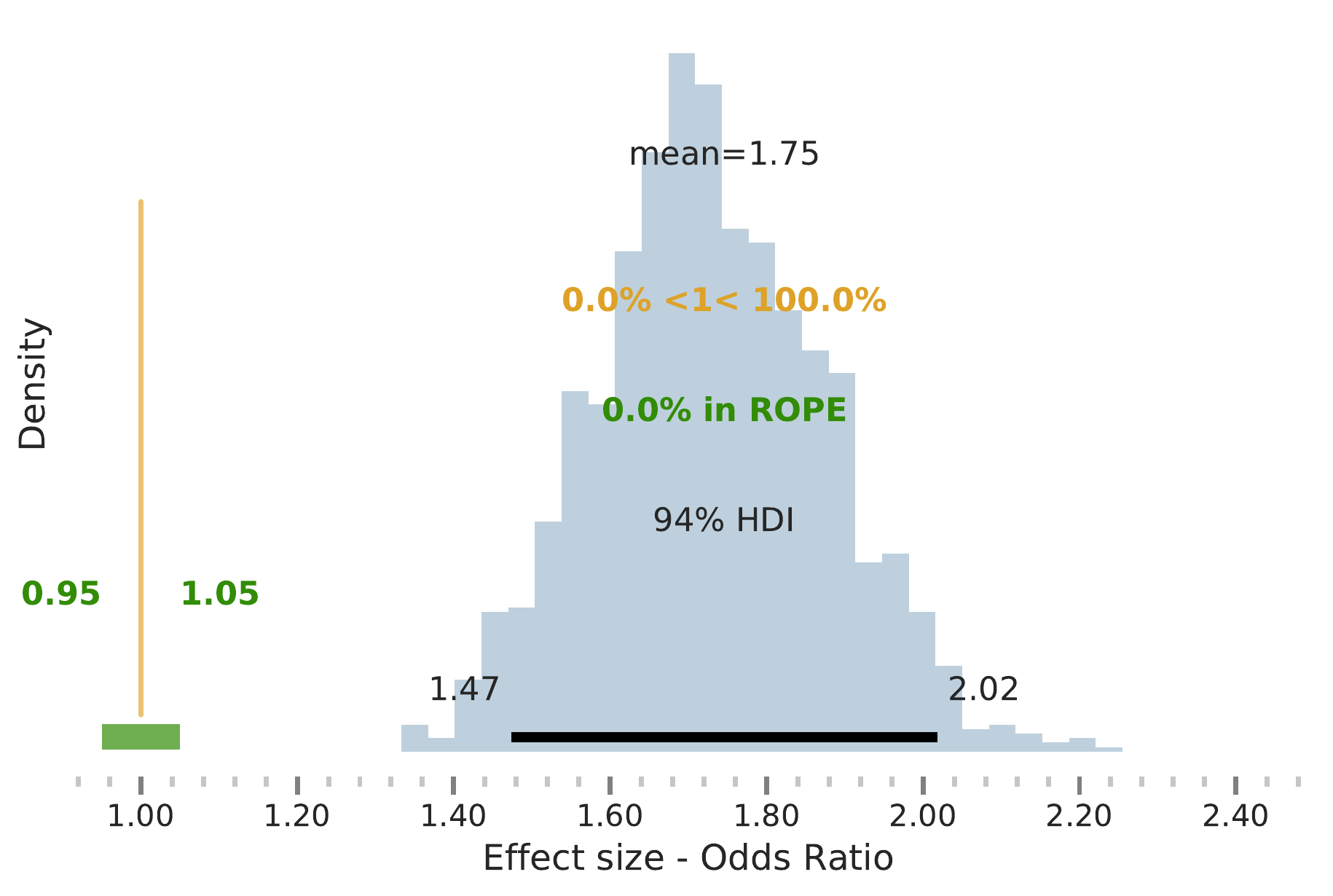}
            \caption[]%
            {{\small Recall, Effect size}} 
            \label{fig:R-effect}
        \end{subfigure}
        \begin{subfigure}[b]{0.30\textwidth}   
            \centering 
            \includegraphics[width=\textwidth]{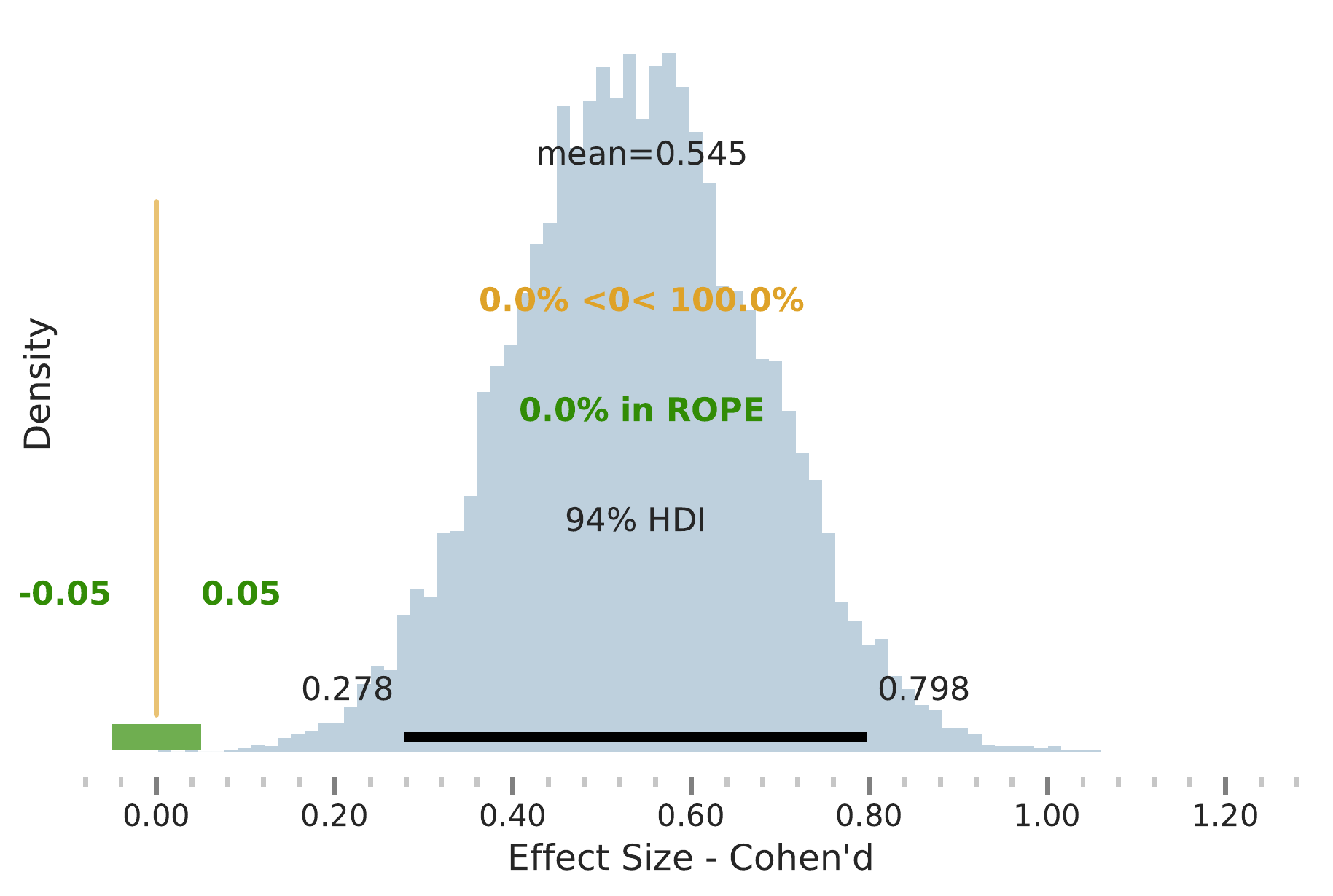}
            \caption[]%
            {{\small Comprehension, Effect size}}   
            \label{fig:C-effect}
        \end{subfigure}
        \caption[] %
        {\small \zx{The first row represents the posterior distribution contrasts between the Chatbot Condition and the Form Condition for the cumulative odds of achieving less than 50\% accuracy in the Recall task and for the means of Comprehension score. The second row shows the effect size distribution, Odds Ratio for Recall and Cohen's d for Comprehension. Plot (a), (b), and (d) show an orange vertical line located at 0 with green bars indicating ROPE. Similarly, plot (c) has an orange vertical line located at 1 and a green ROPE interval. Effect distribution falling into these ROPE regions suggests no difference between conditions or no effect. Note that the x-axis is not the same scale for all plots. \textbf{Main finding:} Compared to the participants in the Form condition, participants in the Chatbot Condition provided more correct answers in the recall task and achieved a higher comprehension score. The differences are both statistically significant.}} 
    \label{fig:reading}
    \end{figure*}

\section{Results}

\zx{Overall, {\tt Rumi} improved participants' consent form reading. Our participants who interacted with {\tt Rumi} could recall more information from the consent form and take more correct actions based on the consent form compared to those in the traditional form-based informed consent process. We also found that in the Chatbot condition, participants perceived themselves as having a more equal power relation with the researcher and offered higher-quality responses in the Dummy Survey Study. Our exploratory path analysis revealed a potential mechanism where the chatbot-based consent method improves response quality by reducing the power gap.}

\subsection{People engaged with Rumi and were satisfied with the experience}
\label{sec:engagement}
\zx{Although our participants spent more time chatting with {\tt Rumi} to complete the informed consent process, they were, in general, satisfied with the Time and Effort (Chatbot: M = 4.72, SD = 1.54; Form: M = 4.48, SD = 1.57) used in completing the informed consent process. They also indicated that they are willing to use such a chatbot for future informed consent experiences (Chatbot: M = 4.21, SD = 1.89; Form: M = 4.03, SD = 1.84). We modeled both measures as ordinal variables and contrasted the posterior distributions of the cumulative likelihoods of a rating of 4 for both conditions. We found that participants’ perceived time and effort and their indicated future use did not differ significantly between the two conditions. The High-Posterior Density Interval (HPDI) \footnote{\zx{The HPDI is the location of 94\% of the posterior density. It is similar to, but different from, the idea of the confidence interval used in non-Bayesian Statistics. In non-Bayesian Statistics, a 94\% confidence interval is informally interpreted as ``with 94\% probability the parameter of interest lies in a specific interval; the tails are of equal width (i.e., 3\%)''; the HPDI is the \textit{densest} interval covering 94\% of the posterior. The HPDI is guaranteed to include the most likely value, but this is not always true for confidence intervals; see McElreath \cite{McElreath2015}. For a more careful definition of the confidence interval, see Hoekstra et al. \cite{Hoekstra2014}.}} for the cumulative odds difference overlapped with the ROPE (Region of Practical Equivalence) \footnote{\zx{Unlike non-Bayesian Statistics, where one can ask, if the two means for two conditions are different $P(\mu_1\neq \mu_2)$, in Bayesian statistics, one asks if the HPDI of the distribution $P(\mu_1-\mu_2)$, the distribution of the difference of the means of the two conditions, excludes an interval where we can consider the two treatments equivalent. This equivalence interval is domain-dependent. A posterior distribution HPDI that lies outside the ROPE is considered a significant result in Bayesian data analysis.}} of $0\pm0.05$ in all cases (Time and Effort: M = 0.05, 94\% HPDI: [0.01, 0.08]; Future Use: M = -0.22, 94\% HPDI: [-0.49, 0.01]), indicating that participants were not significantly more likely to disagree or agree in one condition over the other for both Time and Effort and Future Use.}

\zx{We dug into the participants' transcripts to further understand participants' interaction patterns. We found our participants engaged actively with {\tt Rumi}. Our participants raised a total of 449 questions (M = 3.77, SD = 2.56), and {\tt Rumi} answered 389 (85.97\%) of them. We identified four major categories of questions, {\tt Rumi}'s capability(12.69\%), research team information (11.58\%), study information (56.15\%), and side-talking (19.59\%). Our participants ask about what {\tt Rumi} can do and what questions could {\tt Rumi} answer (e.g., \textit{``What do you know?''}[P107]). Another type of question is about the research team (e.g., \textit{``Who is [Researcher Name]?''}[P57]). Through those questions, the participants could learn more about the research team to start rapport building. Unsurprisingly, people asked the most questions regarding the study itself. Specifically, our participants asked questions about the study procedure (45.23\%; e.g., \textit{``What do I do after the survey?''}[P41]), risks (28.17\%; e.g., \textit{``Will my information be safe?''}[P83]), compensation details (12.30\%; e.g., \textit{``What will I get after this?''}[P94]), and general information about the study (14.30\%; e.g., study purpose, survey topic, etc.). Interestingly, some participants started side-talking with {\tt Rumi}, such as \textit{``How's your day?''}[P43] or \textit{``I didn't sleep well yesterday. Do you sleep?''}[P67], which suggests an even higher engagement. Those questions indicate our participants were willing to spend the effort interacting with {\tt Rumi}. And {\tt Rumi} helped our participant to clarify important information regarding the consent form.}

Prior work suggests that introducing interactivity often creates user burdens which may deter user experience, especially for consent form reading where the required time and effort is one major roadblock \cite{flory2004interventions,de2020implementation}. Our results indicate that people are willing to actively engage with {\tt Rumi} to complete the informed consent process, even though it took a longer time than normal online informed consent. We believe scale and speed should not be the only value in the informed consent process. Sacrificing speed and scale for a more engaging experience and grabbing people's attention, especially in this high-stake scenario, is important to consider.  

\subsection{Rumi lead to better consent form reading}
 \begin{figure*}[t]
        \begin{subfigure}[b]{0.30\textwidth}
            \centering
            \includegraphics[width=\textwidth]{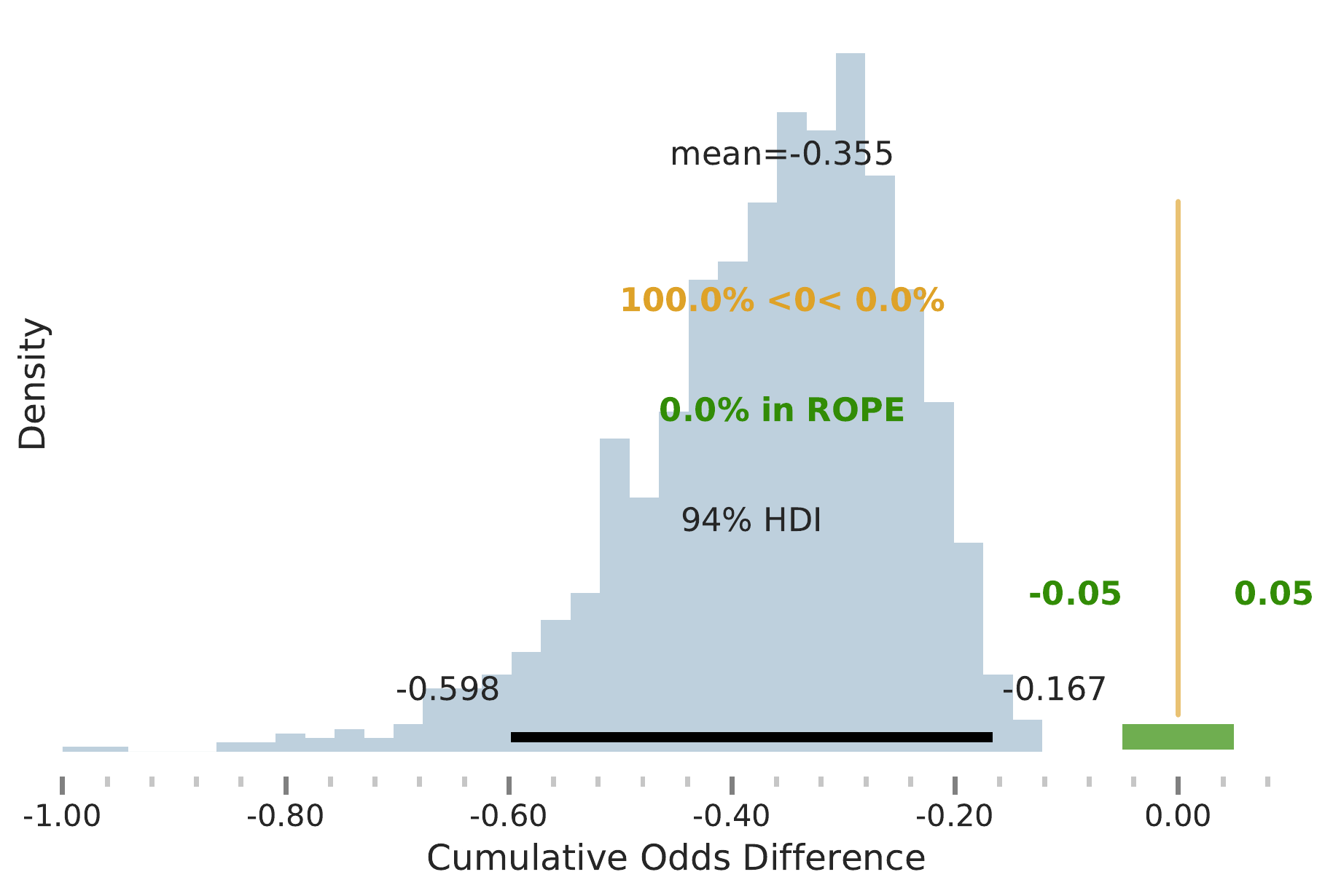}
            \caption[]%
            {{\small Partnership, Contrast}}    
            \label{fig:PN-difference}
        \end{subfigure}
        \hfill
        \begin{subfigure}[b]{0.30\textwidth}  
            \centering 
            \includegraphics[width=\textwidth]{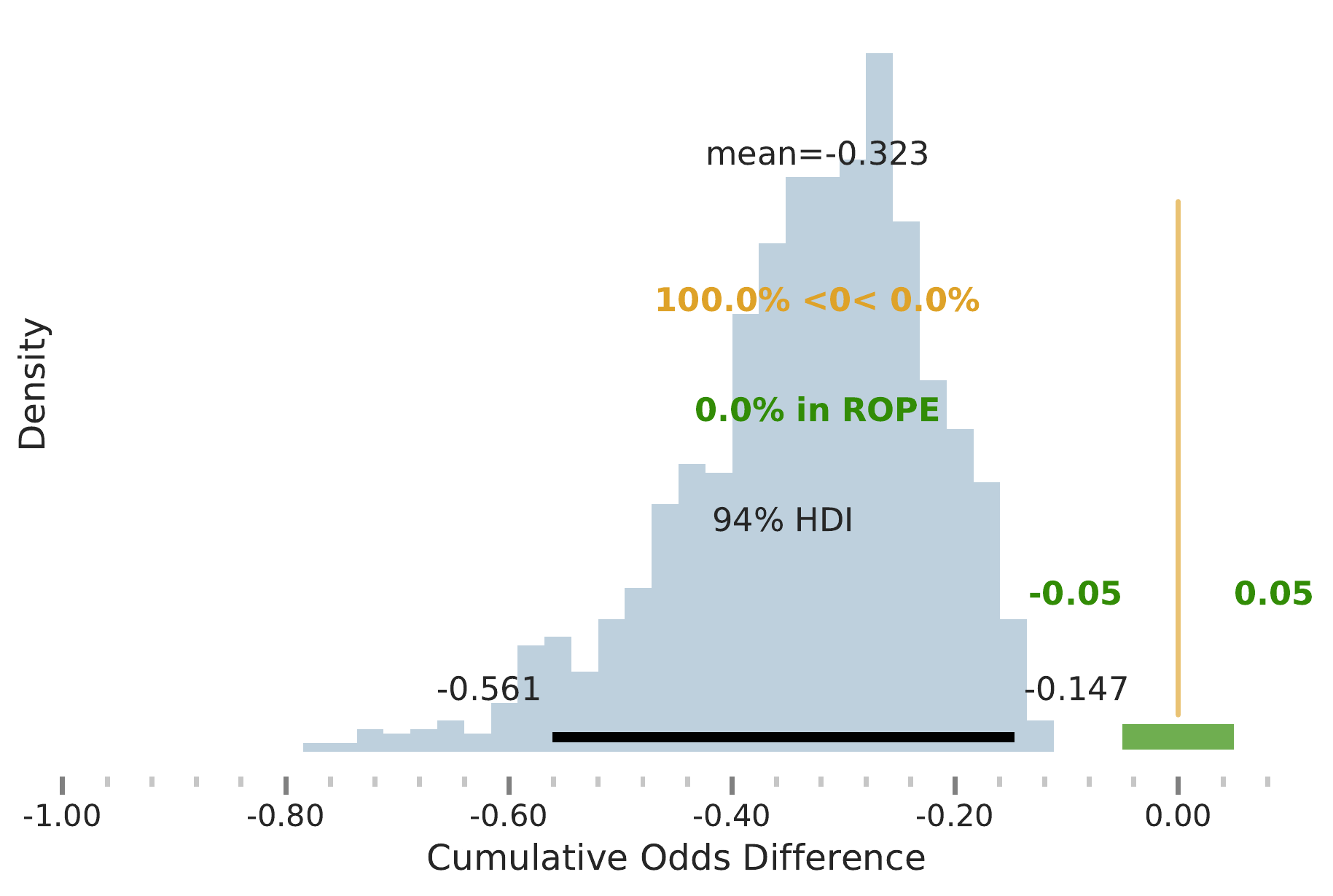}
            \caption[]%
            {{\small Trust, Contrast}}    
            \label{fig:T-difference}
        \end{subfigure}
        \hfill
        \begin{subfigure}[b]{0.30\textwidth}  
            \centering 
            \includegraphics[width=\textwidth]{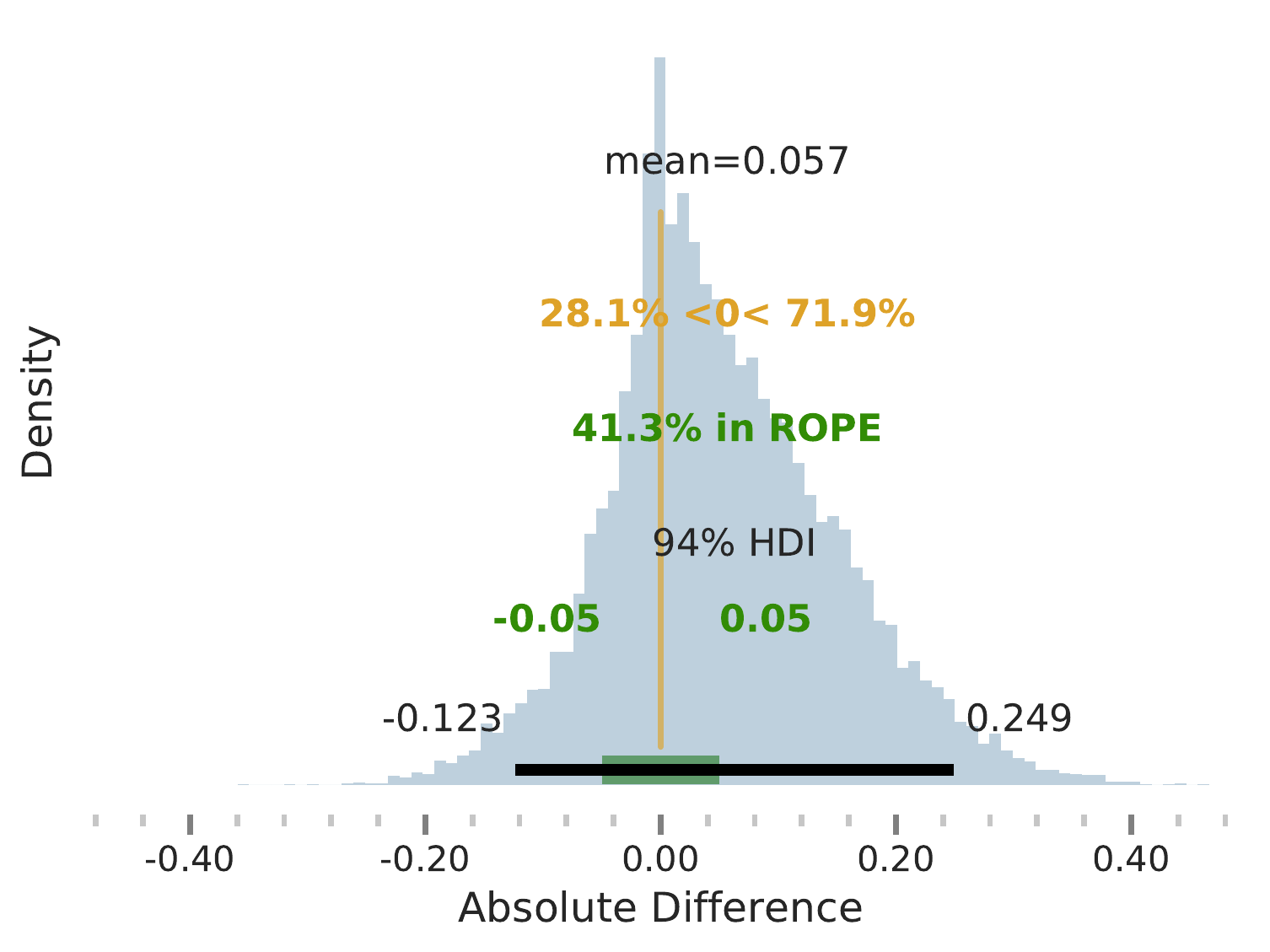}
            \caption[]%
            {{\small Agency and Control, Contrast}}    
            \label{fig:AG-difference}
        \end{subfigure}
        \vskip\baselineskip
        \begin{subfigure}[b]{0.30\textwidth}   
            \centering 
            \includegraphics[width=\textwidth]{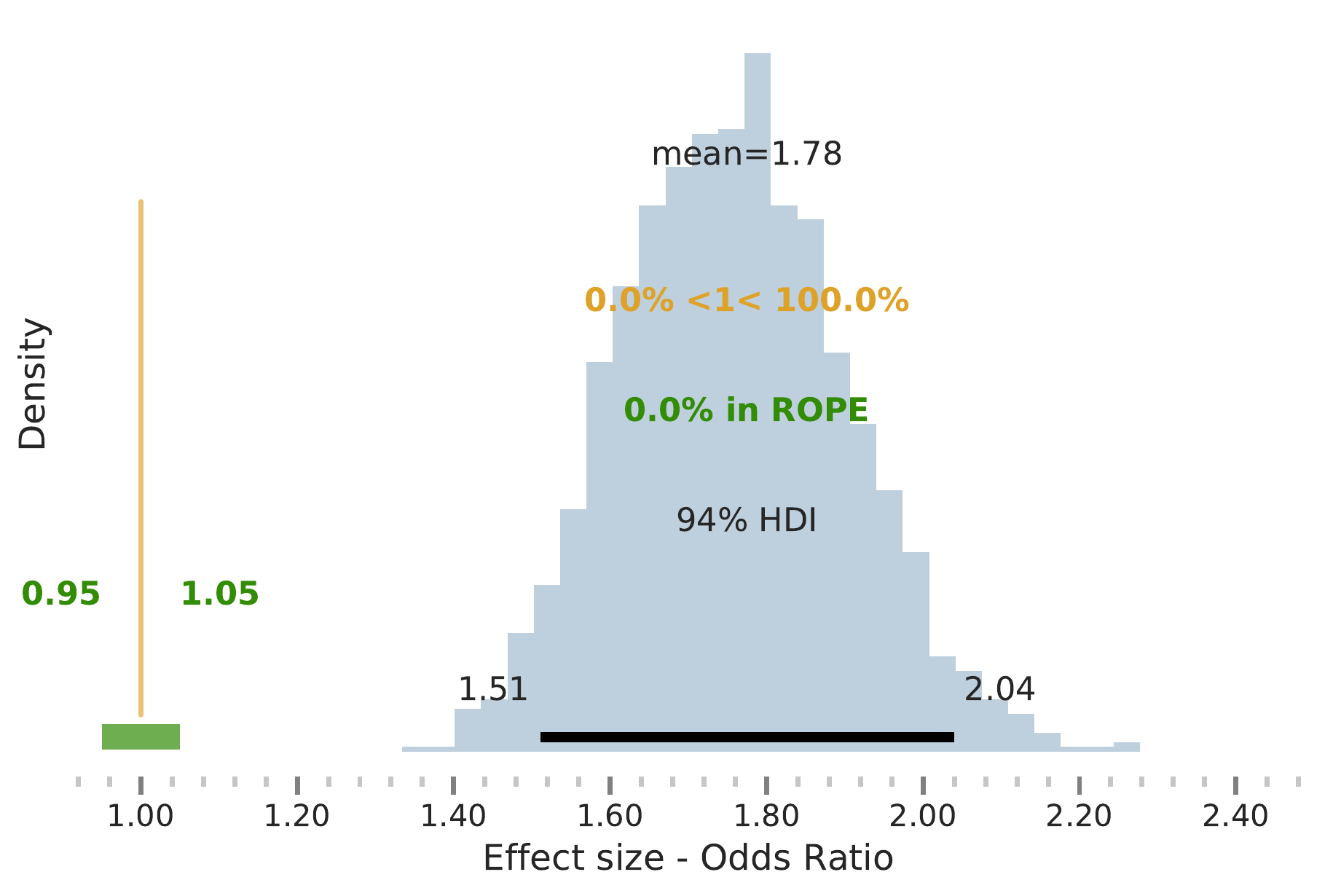}
            \caption[]%
            {{\small Partnership, Effect size}} 
            \label{fig:PN-effect}
        \end{subfigure}
        \hfill
        \begin{subfigure}[b]{0.30\textwidth}   
            \centering 
            \includegraphics[width=\textwidth]{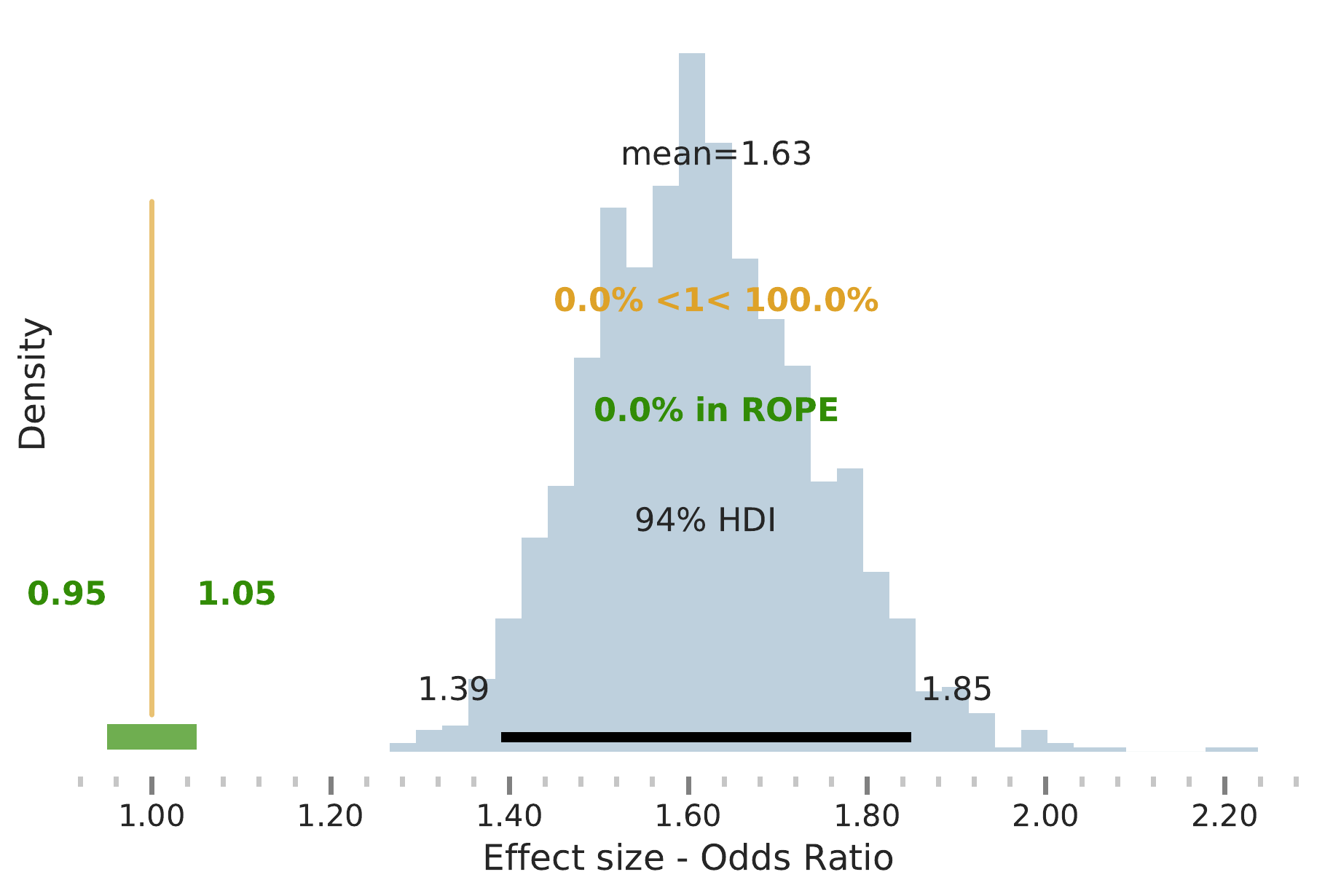}
            \caption[]%
            {{\small Trust, Effect Size}}   
            \label{fig:T-effect}
        \end{subfigure}
        \hfill
        \begin{subfigure}[b]{0.30\textwidth}   
            \centering 
            \includegraphics[width=\textwidth]{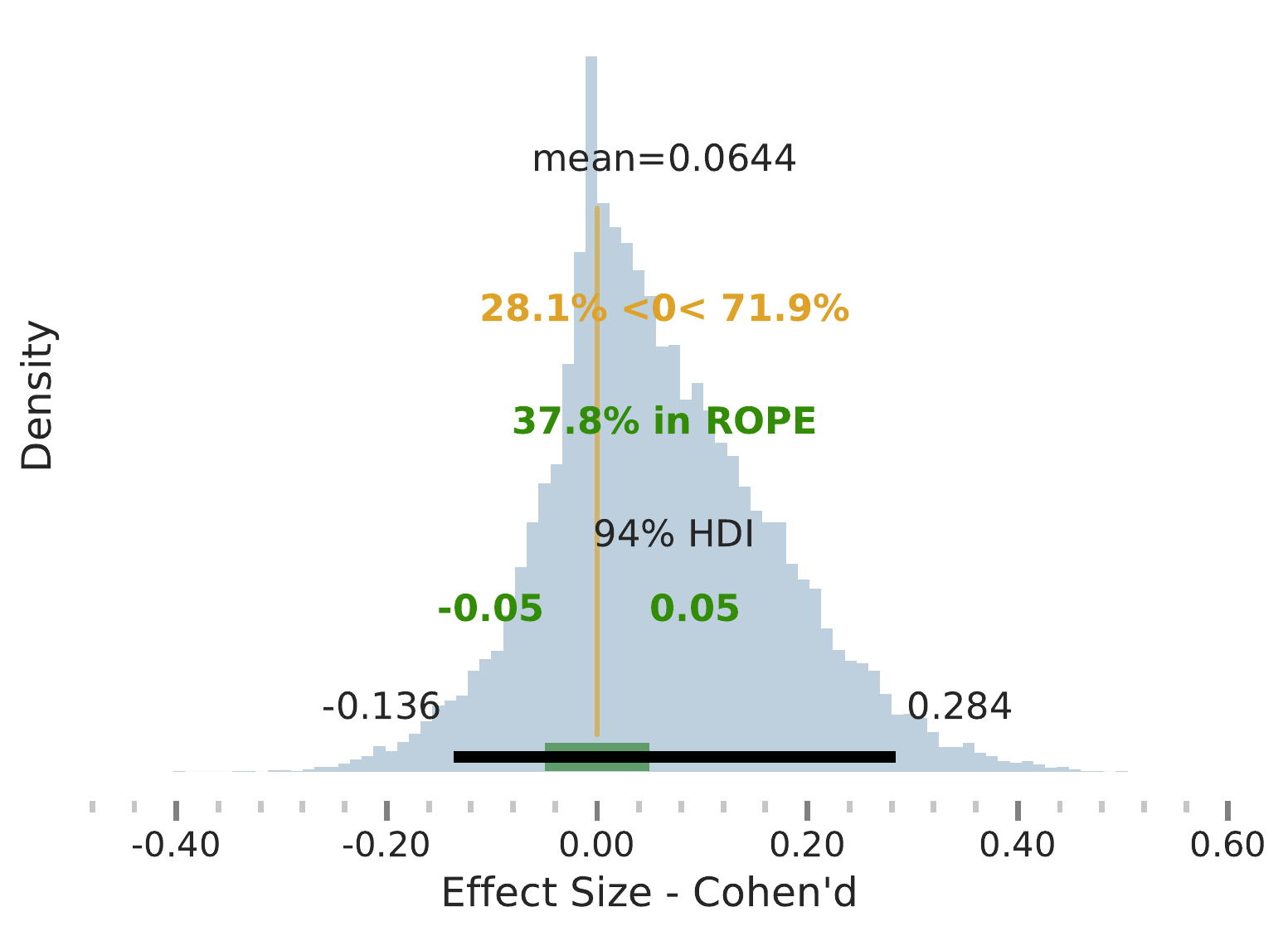}
            \caption[]%
            {{\small Agency and Control, Effect Size}}   
            \label{fig:AG-effect}
        \end{subfigure}
        
        \caption[] %
        {\small \zx{The first row represents the posterior distribution contrasts between the Chatbot Condition and the Form Condition for the cumulative odds of observing a neutral or lower rating in Partnership and Trust and for the means of Agency and Control. The second row shows the effect size distribution, Odds Ratio for Partnership and Trust and Cohen's d for Agency and Control. Plots (a), (b), (c), (f) show an orange vertical line located at 0 with green bars indicating ROPE. Similarly, plots (d) and (e) have an orange vertical line located at 1 and a green ROPE interval. Effect distribution falling into these ROPE regions suggests no difference between conditions or no effect. Note that the x-axis is not the same scale for all plots. \textbf{Main finding:} Compared to the participants in the Form Condition, participants who interacted with {\tt Rumi} perceived their relationship with the researcher more like a partnership and trusted the researcher more. The difference is significant. Although participants in the Chatbot Condition reported higher perceived agency and control, the difference is not statistically significant.}} 
        \label{fig:mean and std of nets}
    \label{fig:pwr}
    \end{figure*}

Overall, similar to prior work \cite{douglas2021some, pedersen2011undergraduate, perrault2019concise}, the consent form reading is poor. In terms of recall, only 26 participants in both conditions in total recalled both ``orange'' and ``blue'' correctly from the color phrase list. \zx{On average, people scored 0.63 out of 2 (SD = 0.64) on the recall task and scored 53.22\% (SD = 24\%) on the comprehension task.} 

\zx{Compared to the static form, going through the consent form with {\tt Rumi} leads to significant improvement in consent form reading. Participants who interacted with {\tt Rumi} recalled more correct color words from the consent form (Chatbot: M = 0.76, SD = 0.71; Form: M = 0.51, SD = 0.54). We modeled Recall as an ordinal variable and estimated the posterior distributions of the cumulative likelihoods of getting 50\% of the total score (1 out of 2). The HPDI for the cumulative odds difference excluded the ROPE of $0\pm0.05$ (M = -3.84, 94\% HPDI: [-6.26, -1.55]). The estimated effect size (Odds Ratio) is small (M = 1.75, 94\% HPDI: [1.47, 2.02], excluding the ROPE of $0\pm0.05$; See Fig \ref{fig:reading}). The result indicates our participants could recall better with {\tt Rumi}.}

\zx{Participants in the Chatbot Condition also better comprehended the consent form (Chatbot: M = 61\%, SD = 22\%; Form: M = 46\%, SD = 23\%). We estimated the posterior distributions of the mean for the Chatbot Condition and the Form Condition. We found the difference between the Chatbot Condition (M = 51\%, 94\% HPDI: [39\%, 62\%]) and the Form Condition (M = 51\%, 94\% HPDI: [27\%, 48\%]) is statistically significant (excluding ROPE $0\pm0.05$), with a medium to large effect size (Cohen's d) (M = 0.55, 94\% HPDI: [0.28, 0.80], excluding a ROPE of $0\pm0.05$; See Fig \ref{fig:reading}).} This suggests that the participants understood the content and could take better actions according to the consent form to protect their rights. 

The results answer RQ1 clearly. The chatbot-driven informed consent process improves consent form reading in terms of both recalling information from the consent form and comprehending its content. Two factors may play a role in the observed improvement. First, the participant may be better engaged. The interactive features of {\tt Rumi} were designed to simulate an in-person experience in which the research assistant actively engaged with the participant. Many of our study participants enjoyed this human-like experience and commented \textit{``the bot is very friendly.''}[P61] and \textit{``i liked how the bot talks to me.''}[P58]. And participants appreciated that {\tt Rumi} went through the consent form with them, \textit{``It was easier and nicer to read the consent form with the bot using texts other than a wall of text. Thank you :D''}[P15]. As we showed in Sec. \ref{sec:engagement}, in the Chatbot Condition, the participant spent significantly more time during the informed consent process. Although time spent may not always lead to better reading \cite{de2020implementation}, it could suggest higher engagement which plays a key role in reading comprehension \cite{grabe2008reading}. Second, {\tt Rumi} is designed to answer people's questions. A total of 449 questions were raised and {\tt Rumi} answered 85.97\% of them. Our participants appreciated {\tt Rumi}'s ability to answer their questions in real-time, \textit{``It’s pretty cool that the chatbot can answer my questions right away''[P32]} but some participants mentioned that {\tt Rumi} cannot fully understand their questions. Although the chatbot has limited capability, the ability to answer people's questions on the fly might contribute to the improved consent form reading.  

\subsection{Rumi aided participant-researcher relationship}

\zx{The participants who interacted with {\tt Rumi} perceived themselves as having a more equal power relation with the researcher in charge of the study.} Our results indicated that people in the Chatbot Condition trust the researcher more (Chatbot: M = 5.60, SD = 1.40; Form: M = 4.92, SD = 1.93) and believed their relationship with the researcher is more like a partnership (Partnership) compared to the Form Condition, which indicates a smaller power gap (Chatbot: M = 4.26, SD = 1.70; Form: M = 3.63, SD = 1.74). \zx{Similar to Sec \ref{sec:engagement}, we treated both variables as ordinal and estimated the posterior distributions of the cumulative likelihoods. Again, we constructed the distribution of the difference between the cumulative odds of a rating of 4 (the midpoint of the 7-point Likert scale) in the Chatbot condition and the cumulative odds of the rating in the Form condition. Based on the estimated posterior distributions, we found the differences of the cumulative likelihoods in both measures are statistically significant (Partnership: M = -0.36, 94\% HPDI: [-0.60, -0.17], excluding the ROPE of $0\pm0.05$; Trust: M = -0.67, 94\% HPDI: [-1.14, -0.28], excluding the ROPE of $0\pm0.05$). The odd ratio suggests a small effect size for both measures (Partnership: M = 1.78, 94\% HPDI: [1.51, 2.04], excluding the ROPE of $0\pm0.05$; Trust: M = 1.63, 94\% HPDI: [1.39, 1.85], excluding the ROPE of $0\pm0.05$). }

\zx{However, we did not observe a significant difference in participants' feelings of agency and control after the informed consent process. The participant who interacted with {\tt Rumi} (M = 5.15; SD = 0.78) reports a higher feeling of agency and control, measured by a composite score of positive and negative agency, compared to the Form Condition (M = 5.04; SD = 0.91). Since we measured participants' feelings of agency and control with a composite score of an 11-item scale, we treated the score as a continuous variable and contrasted the mean between the Chatbot Condition and the Form Condition. We estimated the posterior distribution of the mean difference and found the difference is not statistically significant (M = 0.06, 94\% HPDI: [-0.12, 0.25], overlapping the ROPE of $0\pm0.05$).}

 \begin{figure*}[]
        \centering
        \begin{subfigure}[t]{0.30\textwidth}
            \centering
            \includegraphics[width=\textwidth]{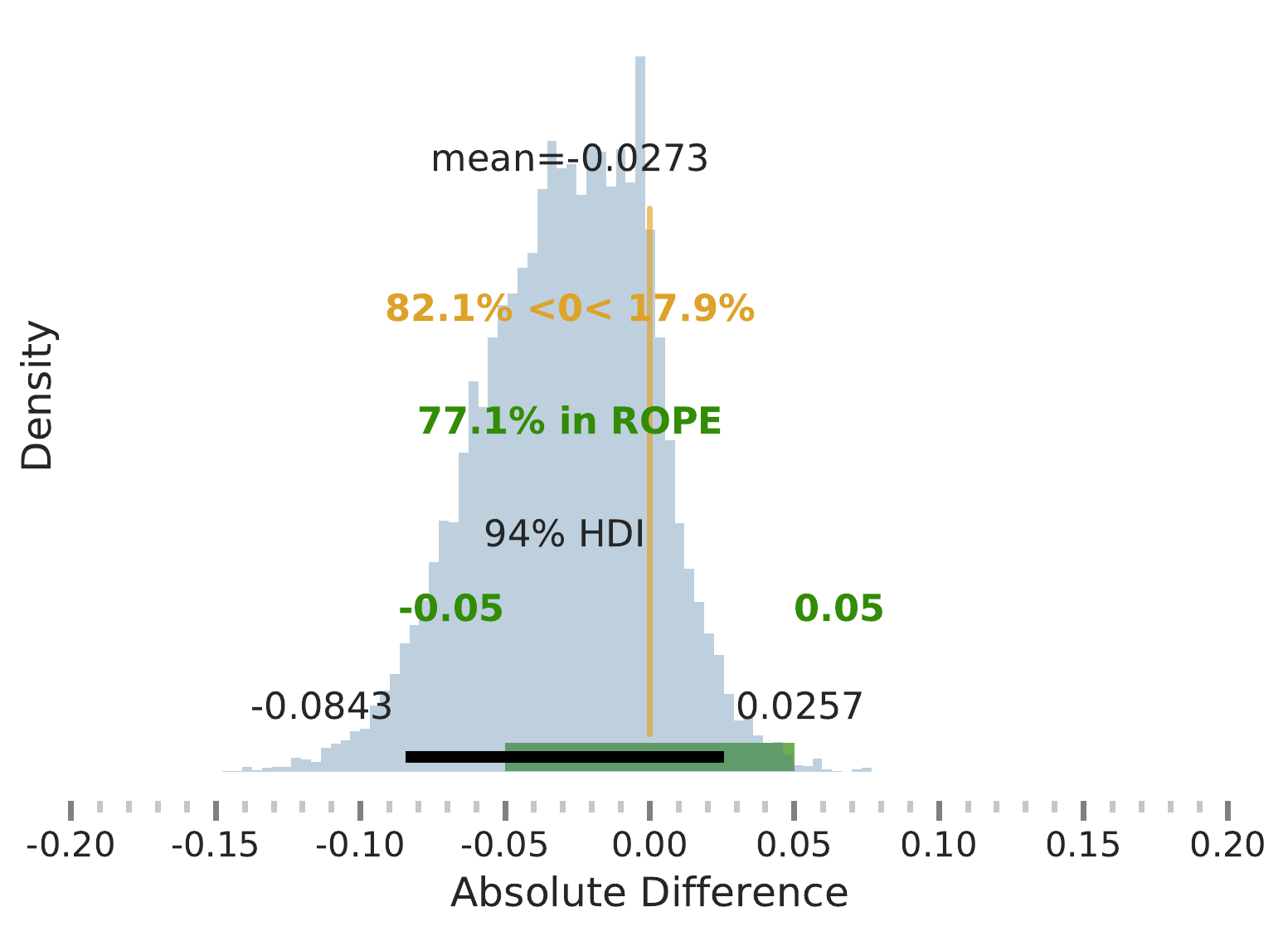}
            \caption[]%
            {{\small Non-differentiation, Contrast}}    
            \label{fig:ND-Contrast}
        \end{subfigure}
        \begin{subfigure}[t]{0.30\textwidth}  
            \centering 
            \includegraphics[width=\textwidth]{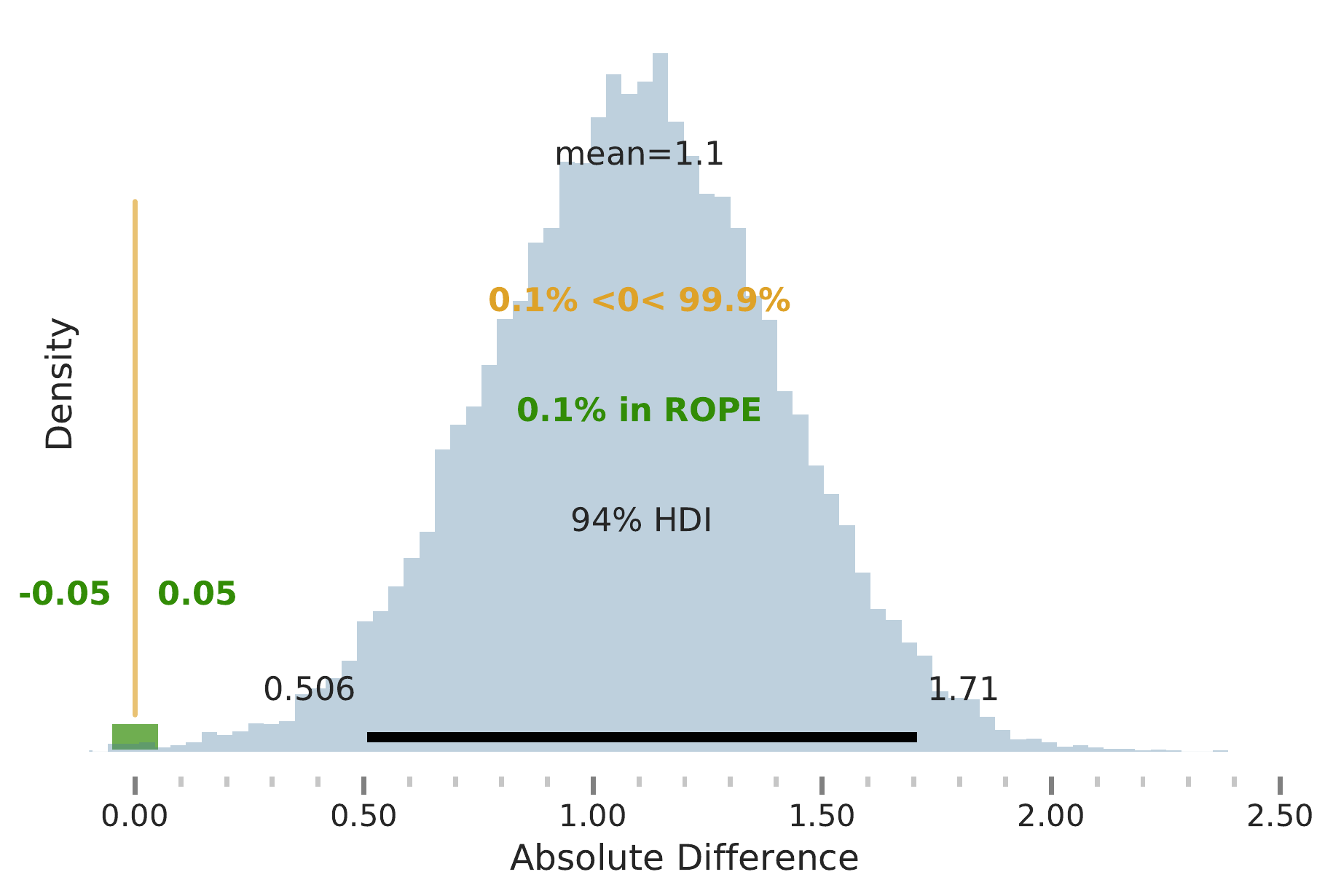}
            \caption[]%
            {{\small RQI, Contrast}}    
            \label{fig:RQI-Contrast}
        \end{subfigure}
        \vskip\baselineskip
        \begin{subfigure}[b]{0.30\textwidth}   
            \centering 
            \includegraphics[width=\textwidth]{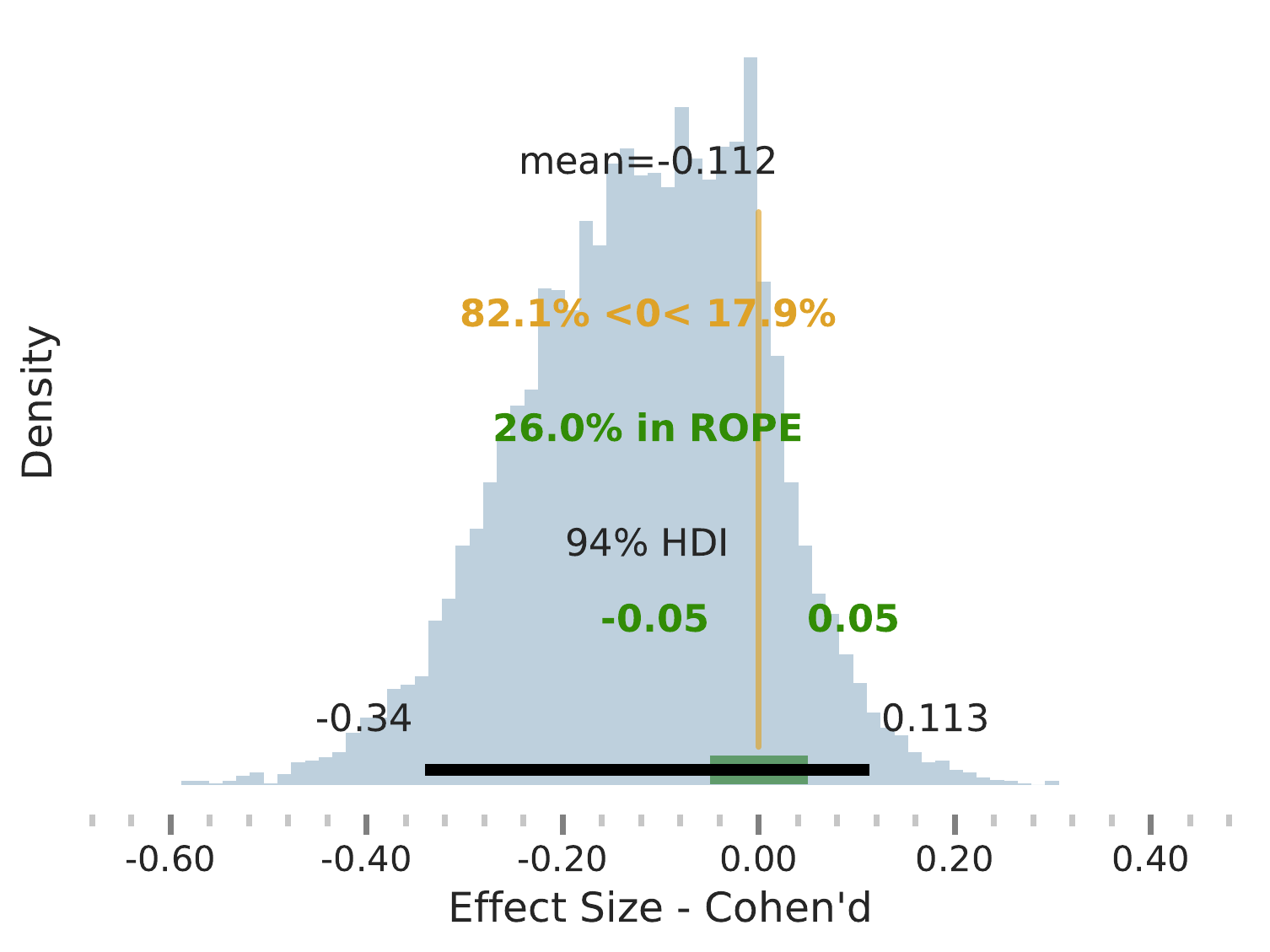}
            \caption[]%
            {{\small  Non-differentiation, Effect size}} 
            \label{fig:ND-Effect}
        \end{subfigure}
        \begin{subfigure}[b]{0.30\textwidth}   
            \centering 
            \includegraphics[width=\textwidth]{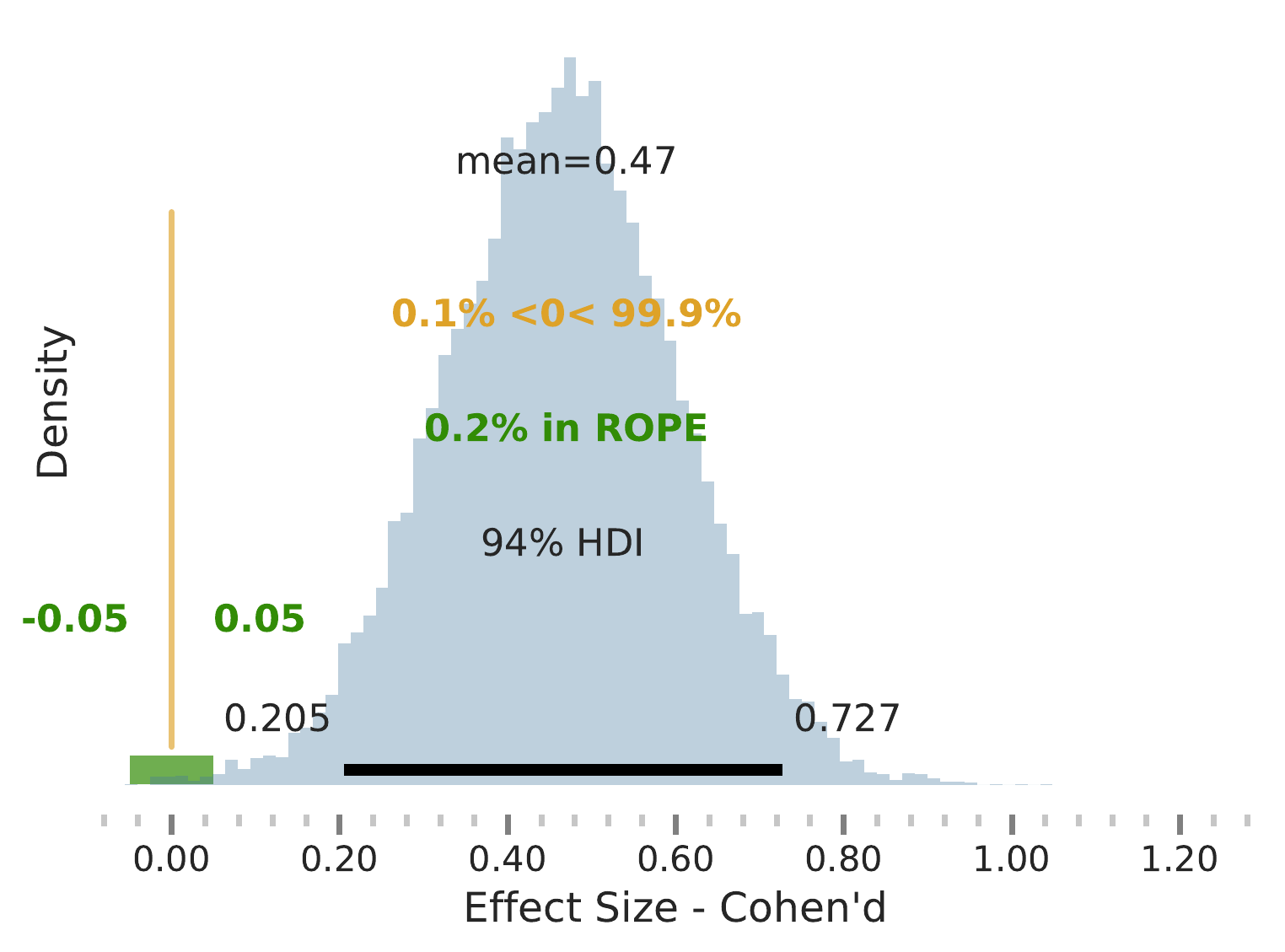}
            \caption[]%
            {{\small RQI, Effect size}}   
            \label{fig:RQI-Effect}
        \end{subfigure}
        \caption[] %
        {\small \zx{The first row represents the posterior distribution contrast of two means between the Chatbot Condition and the Form Condition for the Non-differentiation and Response Quality Index (RQI). The second row shows the effect size distribution, Cohen's d, for both measures based on the posterior distribution. Each plot shows an orange vertical line located at 0 with green bars indicating ROPE. This represents that there was no difference between conditions or no effect. Note that the x-axis is not the same scale for all plots. \textbf{Main finding:} Participants who interacted with {\tt Rumi} displayed less non-differentiation in close-ended questions but the difference is not statistically significant. We also observed participants in the Chatbot Condition contribute higher quality answers to the open-ended questions. The posterior distributions indicate the difference is statistically significant with a medium effect size.}} 
    \label{fig:quality}
    \end{figure*}
    
Going through the online informed consent process with {\tt Rumi} increases the trust between the participant and researcher and closes the power gap. Two potential factors may explain the observed effect. First, as mentioned in \cite{schuck1994rethinking}, a more effective consent form reading could reduce the power gap by bridging the information gap and assuring a voluntary decision. We did observe a significant correlation between participants' consent form reading and their power relation with the researcher (Recall: r(236) = 0.14, p = 0.03; Comprehension: r(236) = 0.21, p < 0.01). The observed difference in the power relation could potentially be attributed to more effective communication. Secondly, the humanness of {\tt Rumi}'s design may help with rapport building between the researcher and the participants, which potentially reduces the power gap \cite{karnieli2009power}. In our study, {\tt Rumi} is framed as a virtual research assistant and represents the research team. As the first interaction between the researcher and the participant, the informed consent process may also serve the role of rapport building beyond informing the study participant. 

\subsection{Rumi lead to better survey response quality}

Participants who interacted with the chatbot provided higher-quality responses to the dummy survey. For choice-based questions, participants in the Chatbot Condition exhibit less survey satisficing behavior (Chatbot: M = 0.42, SD = 0.22; Form: M = 0.47, SD = 0.25). The posterior distributions show the observed difference is not statistically significant (M = - 0.03, 94\% HPDI: [-0.08, 0.03], overlapping a ROPE of $0\pm0.05$). 

For open-ended questions, participants in the Chatbot Condition provided higher-quality responses (Chatbot: M = 5.38, SD = 2.29; Form: M = 4.17, SD = 2.43). \zx{The posterior distribution on the difference of RQI between two conditions shows a mean contrast of 1.1 with HPDI of [0.50, 1.71]. Since the 94\% HPDI lies outside a significant ROPE of $0\pm0.05$, the result implies a significant effect with a medium effect (Cohen's d: M = 0.47, 94\% HPDI: [0.21, 0.73], excluding the ROPE of $0\pm0.05$).}

\begin{figure*}[t]
\includegraphics[width=0.75\textwidth]{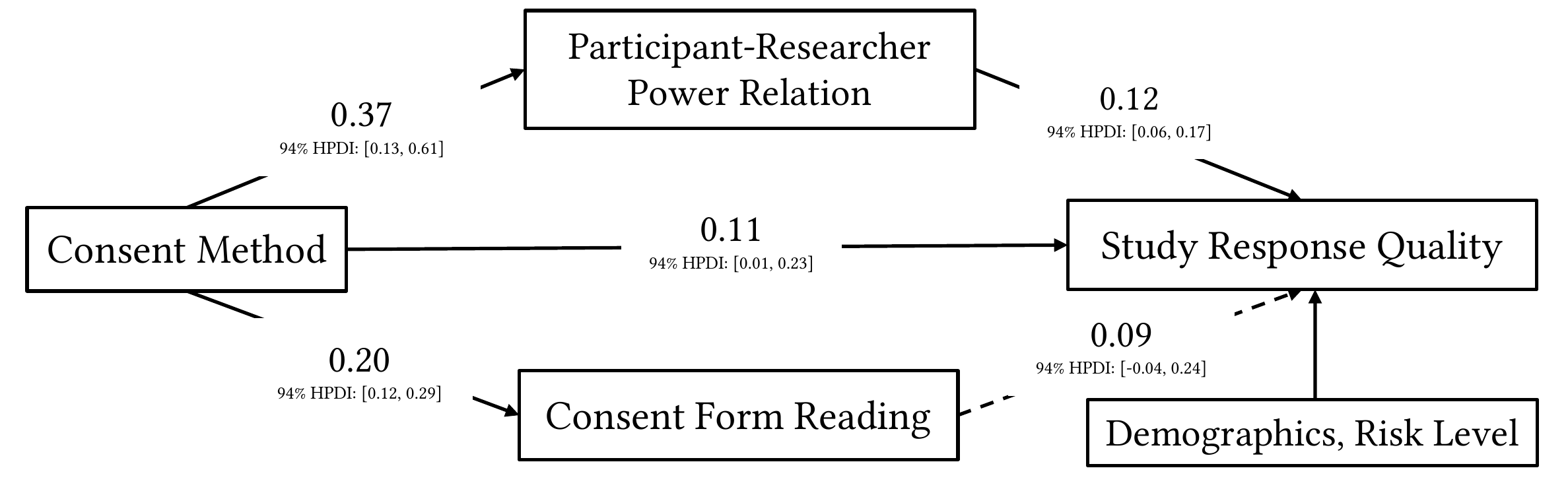}
\centering
\caption{The figure shows a Bayesian Structural Equation Model (SEM) model to understand why chatbot-driven informed consent leads to improved study quality. As suggested by previous literature, we examined two potential pathways, 1) reduced power gap in the researcher-participant relationship, and 2) improved consent form reading. The results indicate a significant partial mediation effect from the reduced power gap.}
\label{fig:sem}
\end{figure*}

The results answered our RQ3. The participants who interacted with the chatbot during the informed consent process contributed significantly higher-quality responses to the open-ended questions. Participants who are in the High-risk condition were also willing to elaborate on their answers. Based on existing work, the observed effect could be due to a more effective consent form reading \cite{del2005informed,douglas2021some} and a reduced power gap in the researcher-participant relationship \cite{cassell1980ethical,karnieli2009power}. We further examined two potential pathways in Sec. \ref{sec:sem}.  

\subsection{Reduced power gap may explain the study response quality improvement}
\label{sec:sem}
\zx{Existing studies suggest that the power gap between the researcher and the participant plays a key role in the participant's engagement and level of self-disclosure \cite{cassell1980ethical,karnieli2009power}. A more equal power dynamic promotes trust and enhances the participant's feeling of agency and control, potentially benefiting the study response quality. On the other hand, better consent form reading may also improve study response quality as the participant understands the study procedure better \cite{del2005informed,douglas2021some}. Therefore, we explore two potential pathways that may mediate the effect of chatbot-driven informed consent on study response quality: through reduced power gap and through improved consent form reading.}

\zx{We built a Structural Equation Model (SEM) model to answer our question. As shown in Fig \ref{fig:sem}, the Consent Method (Chatbot vs. Form) was set to predict the Researcher-Participant Power Relation and Consent Form Reading. The Researcher-Participant Power Relation is a composite score of the Researcher-participant relationship and the participant's perceived agency and control. The Consent Form Reading combines recall and comprehension into one single measure with normalization. The Study Response Quality combines Non-differentiation for choice-based questions and Study Response Quality Index for open-ended questions. Both Researcher-Participant Power Relation and Consent Form Reading were set to predict the Study Response Quality, which is also predicted by the Consent Method. We believe that the Researcher-Participant Power Relation predicts Study Response Quality because the power dynamics influence trust and autonomy, two key factors in self-disclosure \cite{cassell1980ethical,karnieli2009power}. We believe that the Consent Form Reading predicts Study Response Quality by how much the participant understands and follows the study instruction. Whether the Consent Method predicts Study Response Quality is our main question; the other two pathways are paths that mediate the effect of the Consent Method on Response Quality. Like all other models, we controlled participants' demographics and risk levels. All variables were treated as manifest variables and modeled as Normal Distributions. We used Bayesian inference to fit the proposed SEM model with the No-U-Turn Sampler \footnote{https://cran.r-project.org/web/packages/blavaan/}. The model fits perfectly, with posterior predictive p-value = 0.48; BRMSEA = 0.01, 94\% HPDI: [0.00, 0.06]; BMc = 0.99, 94\% HPDI: [0.98, 1.00]; adjB$\hat{\Gamma}$ = 0.99, 94\% HPDI: [0.96, 1.00].}

\zx{Results indicated a significant total effect of the Consent Method on the Study Response Quality ($\beta$ = 0.18, 94\% HPDI: [0.08, 0.29], excluding the ROPE of $0\pm0.05$). Notably, as shown in Fig \ref{fig:sem}, there was an indirect path from the Consent Method to the Study Response Quality via Participant-Researcher Power Relation ($\beta$  = 0.12, 94\% HPDI: [0.06,0.17], excluding the ROPE of $0\pm0.05$). This path reduces the total effect by 39\%. However, the direct effect remained significant, reflecting only partial mediation. Meanwhile, we did not observe a significant path from the Consent Method to the Study Response Quality via the Consent Form Reading ($\beta$  = 0.09, 94\% HPDI: [-0.04, 0.24], overlapping the ROPE of $0\pm0.05$).}

The Bayesian SEM model shows that the potential reason that chatbot-driven informed consent improved later study response quality may be the reduced power asymmetry between the researcher and the participant. This finding aligns with prior studies that as one of the early interactions between the participant and the researcher, the informed consent process could bridge the information gap and shape the researcher-participant power relation, which may ultimately benefit study response quality and data richness \cite{cassell1980ethical,karnieli2009power,schuck1994rethinking}. However, the model does not indicate a potential pathway from the improved consent form reading. We believe there are two potential explanations. First, our study procedure is straightforward. An improved consent form reading may not have a strong effect in preparing the participants for the later study. Second, our study is low-stake compared to medical trials. Existing studies' findings on clinical trials may not generalize \cite{del2005informed}.    

\section{Discussion}
In this study, we found that a chatbot-driven informed consent process could effectively improve consent form reading, reduce the power gap, and ultimately benefit study response quality. In this section, we discuss design implications for a more effective informed consent experience with conversational AI.

\subsection{Personalized Chatbot-driven Informed Consent Experience}
In a turn-by-turn conversation, a chatbot can ask questions about participants' experiences and preferences to initiate personalization \cite{zhou2019trusting}. In our study, we found our participants enjoyed the feeling of a personalized experience. For example, one participant commented, ``\textit{Never seen a bot like this. I like the feeling {\tt Rumi} is talking to ME.''~[P11]}. Given this opportunity, we should consider incorporating personalization into the chatbot-driven informed consent experience. However, we need to be extremely cautious about unwanted persuasive effects \cite{tam2005web} that violate the principle of voluntary participation when introducing personalization into online informed consent. The participation decision should be fully voluntary. The goal of personalization should be to facilitate the understanding of the consent form, not to \emph{nudge} the participant to participate. 

\zx{An AI-powered chatbot could highlight important content based on the participant's previous experiences with consent forms. On the one hand, existing studies suggest that more experienced study participants spent less time reading the consent form and often missed important information \cite{ripley2018uninformed} because they tended to assume all consent forms were similar. Such incorrect assumptions exposed participants to unwanted risks. In this case, their pre-existing experience hindered consent form reading. On the other hand, regulations, such as General Data Protection Regulation (GDPR), often require a consent form to contain information that a participant may already be familiar with. In this case, a chatbot could summarize those materials to help participants better allocate their attention to new content. Therefore, it will be useful for the chatbot to learn about participants' prior informed consent experiences, analyze the new consent form, summarize contents a participant may be familiar with, and highlight new content to ensure a thorough read. An interesting idea to explore is a centralized chatbot that helps participants to manage all consent forms while preserving anonymity from prior study experiences.}      
Although the IRB requires the researcher to draft the consent form in layman's language, some study procedures are indeed complicated. And sometimes certain terminology is necessary for clear communication, especially for high-stake clinical trials. Participants in our study appreciated {\tt Rumi}'s ability to offer clarifications and answer their questions. Therefore, the chatbot should include interactive features to help participants understand the consent form. We could also borrow a thread of chatbot research focused on education where the chatbot helps people study new content and review materials \cite{ruan2019quizbot,colace2018chatbot}. For example, a chatbot could first assess the participant's existing knowledge about the study topic to determine the necessary explanations and ask questions at the end to ensure learning outcomes. 

\zx{However, as time and effort are among the biggest hurdles in consent form reading, the trade-off between the benefits and risks of the interactive features needs proper calibration and consideration. Specifically, we need to consider the time cost a participant will spend on consent form reading. A participant's compensation is often associated with time spent in a study procedure. If the total compensation is fixed, the interactive informed consent procedure will reduce the pay rate. In our study, we compensated participants in the Chatbot condition with an extra bonus at the end to ensure the promised pay rate. Although the research team may need to budget more for each participant, we argue, as the study shows, the improved response quality will ultimately benefit the research results. It saves time for researchers to clean up low-quality data from participants who did not read the study procedure carefully, especially in cases where unattended participants create confounding factors that endanger the study quality \cite{douglas2021some}.}

\subsection{Managing Power Dynamics}
Our results echo prior studies on the role of an effective informed consent process for closing the power gap in the researcher-participant power relation and its benefit in study response quality. We believe that the chatbot could influence researcher-participant relations by adjusting its own power relation with the participant. We could further extend the utility by designing an informed consent chatbot that actively manages the power dynamic.  

From a power relation point of view, one's identity has a strong influence on their relative power over their counterpart \cite{foucault1982subject}. Cassell mentioned that a researcher's chosen identity, e.g., Interviewer, Facilitator, Initiator, Researcher, could change the power dynamic with the study participant \cite{cassell1980ethical}. We could carefully design an informed consent chatbot's identity to suit various contexts. For example, to reduce the power gap, we could design the chatbot as a research partner rather than as a researcher. However, designing a virtual agent's identity is complicated. Many design dimensions, including appearance, language style, etc., need to be carefully considered. Any incoherence may mar the entire experience.

The informed consent process could be considered as a negotiation about information disclosure between the researcher and the participant \cite{karnieli2009power}. The researcher holds the information about the study, and the participants gain the knowledge and experience needed to perform the study. Karnieli-Miller et al. pointed out that such negotiation has the potential to change power relations by giving participants more information \cite{karnieli2009power}. Thus, we should prepare an informed consent chatbot with conversation skills for such negotiation, so that the chatbot could understand participants' requests, clarify their information needs, and actively manage information disclosure about the study. 

We could further empower the participant by considering the ownership of the informed consent chatbot. In this study, the chatbot acted on the researcher's behalf and as a part of the research team. Although in most chatbot use cases the chatbot is owned by the creator instead of the user, a participant-owned informed consent chatbot may provide several benefits. First, the participant will have total authority over the conversation history. In this case, the participant could have a safe space to ask questions without feeling judged. Second, such an informed consent chatbot could become the central hub for all informed consent needs. It could act on behalf of the participant, analyze consent forms based on the participant's preferences, and proactively ask the researcher questions to satisfy the participant's information needs. 

\subsection{Combining Human Expertise with LLMs}
Many of our participants liked {\tt Rumi}'s ability to respond to their questions in real-time with answers grounded in the consent form content. However, creating a chatbot that can accurately answer people's questions, especially in high-stake contexts, is challenging. Due to limited natural language understanding ability, the current Q\&A functionality for most commercial chatbot building platforms relies on a database of handcrafted Q\&A pairs. It is especially time-consuming in the informed consent context as participants' questions are specific to the consent form, further limiting the reusability of a Q\&A database. Although some questions could be reused, for example, an institution may share the same template, and some study procedures could be similar, future studies are needed to design tools to support such a sharing practice.

Large language models (LLMs), like GPT-3, show promise in a new way to build conversational agents to answer people's natural language questions. Although one could use off-the-shelf LLMs with in-context learning to build a chatbot to answer a wide range of domain-specific questions, LLMs sometimes generate non-factual information and have limited capability to memorize a long document~\cite{brown2020language}. Both shortcomings should be avoided in high-stake contexts, e.g., delivering consent forms, as non-factual information could mislead a participant to make an uninformed decision. It is not only an ethical concern but also could lead to severe consequences. For example, a participant who agrees to join a study without full knowledge of the specific study procedures may experience unexpected extreme physical or mental stress.

Therefore, we should consider leveraging LLMs carefully with the above shortcomings in mind. One framework to consider is combining LLMs with human expertise. In our study, we used GPT-3 to augment Q\&A pair generation to empower {\tt Rumi}. To ensure correctness, we acted as a validation layer to check if the GPT-3 generated paraphrased questions and answers were correct and appropriate. The augmented Q\&A database enables {\tt Rumi} to capture more participant questions and delivered more diverse answers. We believe LLMs could facilitate more chatbot development tasks by teaming with human experts. For example, one could use an LLM as a testing tool by generating question sets to identify issues and develop fix. Such method could enable a faster iteration that traditionally relies on bootstrapping conversations on the fly~\cite{xiao2023powering}. Besides using human expertise as a gatekeeper, we should study better LLMs control mechanisms for factual Q\&A. For example, we could leverage a knowledge-driven approach \cite{ge2022should} by parsing a plain text consent form into a structured knowledge graph and using the graph to steer LLMs to generate factual answers that are grounded in the consent form content. Again, given the shortcomings of generative models, we believe a human-in-the-loop framework is preferred to safely take the advantage of generative models for more capable informed consent chatbots. 

In summary, future work should study effective human-in-the-loop frameworks that can support research teams, especially teams without AI expertise and resources, to build and test an informed consent chatbot that consistently delivers factual answers.

\subsection{Future Directions}
\subsubsection{Chatbot as a Virtual Research Assistant}
In addition to prior work that studies the uses of chatbots for research, including for conversational surveys \cite{xiao2020tell} and for ethnographic studies \cite{tallyn2018ethnobot}, we did see a future opportunity to build virtual research assistants that could help researchers manage human-subject studies from beginning to end. A virtual research assistant could help researchers reach a worldwide population (if necessary for the study), engage with the participant and build rapport, deliver the intervention, collect high-quality data, and debrief the participants. Beyond moderating the study process, in our study, we found some participants wanted to communicate with the researcher through the virtual agent, one participant said to {\tt Rumi}, \textit{``let your owner know i like this study.''}[5]. Such a virtual research assistant could be especially helpful for longitudinal field studies where keeping participants engaged and collecting high-quality data over a long period of time is particularly difficult and expensive.

However, creating such a virtual research assistant is challenging. First, repeated interactions with a virtual research assistant at different stages of a study pose new challenges to interaction design. The agent needs to adapt and react to unique individual experiences over time. Second, personalization is a double-edged sword in the context of human subject research. Although it could increase engagement, a highly personalized chatbot could induce unwanted confounding factors due to the inconsistency across participants. Third, mediating the communication between the participants and the researchers requires a new interaction paradigm. The agent needs to mediate the communication and actively engage both parties for study success. Though challenging, such a virtual research assistant could help researchers conduct scalable, robust, and engaging human-subject studies.   

\subsubsection{Online Informed Consent Beyond the Research Context}
Personal data, from personal health information to web browsing histories, has become increasingly valuable. It powers millions of intelligent applications. As the world becomes more connected, personal data becomes a new source of power. As a result, while system builders are eager to collect data, policymakers and users are more cautious about personal data sharing. For example, GDPR \footnote{https://gdpr-info.eu/} regulations explicitly require data collectors to ask for users' consent before collecting and storing any personal data. However, the current practice of data consent online is largely flawed~\cite{barocas2014big}. For example, users often need to go through a lengthy document without any guidance. Therefore, how to empower user giving meaningful and informed consent about their data becomes an emerging challenge. In this study, we found an AI-powered chatbot could deliver effective informed consent in the context of human subject research. In the future, we should explore the potential of such an agent for data-sharing consent in broader contexts. 
  
\subsection{Limitations}
\zx{We recognize several limitations in our approach. First, as the first study of this kind, our main goal was to explore the potential benefits and limitations of the informed consent process driven by an AI-powered chatbot. Through an SEM model, we further explored the potential path that may explain our observed effect; namely, the chatbot-driven informed consent process improved the study response quality by altering the power relationship. However, due to the exploratory nature of this study, our study could not infer strong causal relationships. Future confirmatory studies are needed to confirm the observed effects and explain the mechanism.}

Second, the scope of our study design was limited to online studies with surveys. Although we designed three risk levels to simulate studies that collect different types of data, compared to high-stake clinical trials that may involve severe ramifications, the risk of an uninformed decision is lower in our case. Two factors may limit the generalizability of such a design. First, people tend to pay less attention to consent forms for lower stake studies \cite{pedersen2011undergraduate}. We may observe a smaller difference if people were more attentive to the consent form in both conditions. Second, the study procedure was simple and straightforward, e.g., complete a survey. Although it represents the majority of online studies, some studies may include more complicated study procedures, for example, playing a game, where a good understanding of the procedural details may play a stronger effect in the later study. Thus, we need to study chatbot-driven informed consent under various contexts and studies with different levels of complexity.

Third, participants who declined to join the study are missing from our analysis. Even though, in our study design, participants who decided not to join the study were redirected to Section 2 and offered the opportunity to complete the consent form evaluation. However, some participants may have closed the consent form without answering it. Although other factors may play a role, e.g., usability issues, those participants could read the consent form carefully and make an informed not-to-participate decision. The current study design did not include those participants in the analysis. In our study, 26 out of 278 participants opened our consent form (Chatbot Condition: N = 18; Form Condition: N = 8) without completing it. We believe the effect of this potential confounding factor on our results is limited, but a future study is necessary. 

Fourth, our study was designed to investigate the \textit{holistic} effect of using an AI-powered chatbot to lead the informed consent process. However, the design of a chatbot (e.g., language style, name, and appearance) and its capability (e.g., natural language interpretation, question answering, dialogue management) are important design dimensions that may have an effect on the final outcome. As the first step, we aimed to build {\tt Rumi} to deliver the best possible experience. The data collected in this study were inadequate to tease apart and quantify the contribution of each individual design factor. Since each of the interaction features have both benefits and risks \cite{xiao2020tell}, it is valuable to rigorously quantify the contribution of different features. This, however, requires additional, fully controlled experiments that are beyond the scope of the current study.

Lastly, although chatbots are increasingly adopted in our daily lives, from customer service to conversational surveys \cite{grudin2019chatbots}, it is still uncommon to use chatbots to conduct an informed consent process. Used in the first study of its kind, {\tt Rumi} was novel to most participants. Since we could not control for the novelty effect in our current study design, we did not know the impact of novelty factors. While we are planning longitudinal studies to examine the influence of the novelty effect, the novelty effect may wear off, like any novel technology, as chatbots become a norm. In our case, as chatbot-driven informed consent becomes more common, the effect is most likely to wear off, similar to how more studies are using e-consent forms today \cite{de2020implementation}.

\section{Conclusion}
\zx{In this paper, we examine the role of an AI-powered chatbot in improving informed consent online. We built, {\tt Rumi}, an AI-powered chatbot that can greet a participant, go through the consent form section by section, answer the participant's questions, and collect people's consent responses, to simulate an in-person informed consent experience. We designed and conducted a between-subject study that compared {\tt Rumi} with a typical form-based informed consent process in the context of an online survey study about people's social media use to examine the holistic effect of a chatbot in leading an online informed consent process. We found {\tt Rumi} improved consent form reading, promoted a more equal power relationship between the participant and the researcher, and improved the study response quality. Our exploratory path model indicated the improved study response quality may be attributed to the reduced power gap by the chatbot-driven informed consent process. Given our study results and the simplicity of creating such a chatbot, our work suggests a new and promising method for conducting effective online informed consent. As chatbots become more popular, our results also present important design implications for creating more effective informed consent chatbots. }

\begin{acks}
We would like to thank Heng Ji, Brent W. Roberts, Yu Xiong, Michelle X. Zhou, and the anonymous reviewers for their thoughtful comments and constructive feedback on this work. Tiffany Wenting Li was supported by Google PhD Fellowship Program.
\end{acks}
\bibliographystyle{ACM-Reference-Format}
\bibliography{interactiveconsent}

\end{document}